\documentclass[aps,prl,reprint,superscriptaddress,10pt]{revtex4-2}

\usepackage{graphicx,amsmath,amssymb,dcolumn,physics,color,bm}

\RequirePackage{xcolor}
\definecolor{midgreen}{rgb}{0.17, 0.63, 0.17} %tab:green
\definecolor{darkblue}{rgb}{0.12, 0.47, 0.71} %tab:blue
\definecolor{purple}{rgb}{0.58, 0.40, 0.74} %tab:purple
\usepackage{xr}
\RequirePackage{hyperref}
\hypersetup{linktocpage=true, colorlinks=true, citecolor={midgreen}, linkcolor={purple}, urlcolor={darkblue} }

\newcommand{\x}{\bm{x}}

\newcommand{\diffusivity}{\bm{D}}
\newcommand{\de}{\mathrm{d}}

\newcommand{\xtraj}{\{\bm{x}\}_{-\infty}^t}
\newcommand{\dxc}{\mathrm{d}\bm{x}^c}
\newcommand{\dxctraj}{\{\mathrm{d}\bm{x}^c\}_{-\infty}^t}
\newcommand{\y}{\bm{y}}
\newcommand{\z}{\bm{z}}
\newcommand{\type}{Letter }

\newcommand{\maybespace}{\;\;\;} 
\newcommand{\physics}{Department of Physics and Quantitative Biology Institute, Yale University, New Haven, CT, USA}

\newcommand{\mcdb}{Molecular, Cellular, and
Developmental Biology, Yale University, New Haven, CT, USA}

\makeatletter
\newcommand*{\addFileDependency}[1]{% argument=file name and extension
  \typeout{(#1)}
  \@addtofilelist{#1}
  \IfFileExists{#1}{}{\typeout{No file #1.}}
}
\makeatother

\newcommand*{\myexternaldocument}[1]{%
    \externaldocument{#1}%
    \addFileDependency{#1.tex}%
    \addFileDependency{#1.aux}%
}

\myexternaldocument{supp}
\begin{document}

\title{Passive Entropy Rate Bounds the Information Required for Feedback Control}

\author{Matthew P. Leighton}
\email{matthew.leighton@yale.edu}
\affiliation{\physics}
\author{Jose M. Betancourt} 
\affiliation{\physics}
\author{Thierry Emonet} 
\affiliation{\physics}
\affiliation{\mcdb}
\author{Benjamin B. Machta}
\affiliation{\physics}
\author{Michael C. Abbott}
\affiliation{\physics}

\begin{abstract}
Living organisms across scales, along with many engineered devices, must control noisy systems with limited information to maintain a desired steady state.
Here we ask how the performance of a feedback controller is limited by the rate at which information flows from sensors to control actions.
We demonstrate that the minimum information rate required is the rate at which the passive dynamics increase the system's entropy.
This general result allows us to quantify the information required to achieve a given target performance for a wide-ranging set of control problems.
Next, we ask what explicit control protocol maximizes performance for a given information rate. We demonstrate that in many cases the best control strategy is to probabilistically time-reverse the uncontrolled passive dynamics.
Implementing this protocol approaches or saturates our bound across a wide range of applications.
We showcase these results through applications to stochastic bit control, nonlinear particle localization, microbial navigation, and experimentally realized information engines. 
Our findings pave the way for the study of feedback control in any settings where limited information is a key constraint, suggest principled explanations for microbial behavior, and provide new insights into fundamental questions regarding the thermodynamic value of information.
\end{abstract}

\date{\today}

\maketitle

%%%%%%%%%%%%%%%%%%%%%%
%\section{Introduction}
The ability to optimally control a physical system is fundamentally limited by the availability of both energy and information~\cite{bechhoefertextbook}. Determining control protocols which minimize energetic costs has been a central focus of stochastic thermodynamics~\cite{schmiedl2007optimal,sivak2012thermodynamic,machta2015dissipation,horowitz2017minimum,horowitz2017information,blaber2023optimal}. Information-limited control has been less thoroughly explored, despite its importance in contexts such as single-cell decision-making~\cite{mattingly2021escherichia}, evolution~\cite{nourmohammad2021optimal}, and nanoscale engineering~\cite{saha2022bayesian}. 

The Data Rate Theorem~\cite{tatikonda2004control,nair2004topological,nair2007feedback,garcia2021ergodicity} provides a widely used bound on the information required to stabilize an unstable dynamical system.
For linear deterministic systems, stabilization requires an information rate greater than the sum of positive eigenvalues of the system. While this threshold is useful, it does not tell us how much information is required to achieve higher-precision control, and has not been generalized to the stochastic dynamical systems widespread throughout biology. Related work on information-limited control more generally under various simplifying assumptions~\cite{touchette2000information,touchette2004information,todorov2009efficient,tishby2010information}, and specific applications such as molecular fluctuation suppression~\cite{lestas2010fundamental} and navigation~\cite{mattingly2021escherichia,betancourt2026discrete}, have not yet led to widely applicable information bounds for general control problems. 

Information-limited control has also been studied in the context of information thermodynamics~\cite{Landauer1961irreversibility,still2012thermodynamics,Parrondo2015_Thermodynamics}. Generalizations of fluctuation theorems~\cite{sagawa2010generalized,horowitz2010nonequilibrium,ponmurugan2010generalized,sagawa2012nonequilibrium} and the framework of bipartite thermodynamics~\cite{horowitz2014thermodynamics,horowitz2014second,ehrich2023energy,ehrich2023energetic,leighton2024information,leighton2025flow} have deepened our understanding of connections between energetic costs and information flows between subsystems. Work in this area has, however, focused on systems with a clear thermodynamic interpretation, which significantly limits applicability to complex biological systems.

\begin{figure}[t]
    \centering
    \includegraphics[width=\linewidth]{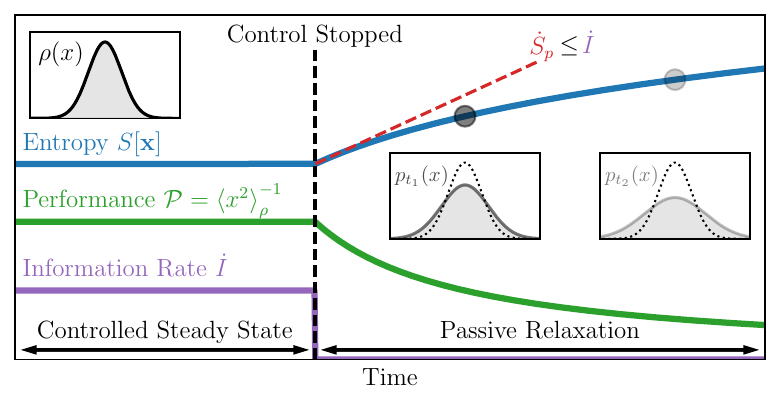}
    \caption{Schematic illustrating our main result. A Brownian particle in one dimension is controlled to remain near $x=0$, producing a narrow steady-state distribution $\rho(x)$ and high performance $\mathcal{P} = 1/\langle x^2\rangle_\rho$. If the control is stopped, the system relaxes under the passive dynamics, with decreasing performance and increasing entropy $S[x]$. The rate at which the entropy increases at the instant the control is stopped, $\dot{S}_p$, bounds the information rate $\dot{I}$ required for control.
    \vspace{-3mm}
    }
    \label{fig:schematic}
\end{figure}

In this Letter, we show that steady-state control performance is limited by the rate of information flow from the state of the system to the controller. This result constitutes a powerful tool for quantifying how much information is required to achieve a given degree of control performance. For a broad class of systems, we derive an optimal control protocol which saturates the bound by probabilistically time-reversing the dynamics of the uncontrolled system. We demonstrate the power of our results through applications to biology, engineering, and information thermodynamics.

%%%%%%%%%%%%%%%%%%%%%%
\emph{Information Limit for Feedback Control.---}
Consider a stochastic system with state $\x$. In the absence of control, it follows Markovian dynamics that we refer to as ``passive", indicated by superscript $p$. The passive dynamics produce an update $\de\x^p_t = \x_{t+\de t}-\x_t$ over time $\de t$ distributed according to the conditional distribution $p(\de \x^p_t|\x_t)$. The passive update might generally include deterministic drift, diffusion, or discrete jumps. 

We seek to control this system so that the state $\x$ obeys a steady-state distribution $\rho(\x)$. The true goal will generally not be to control the full distribution $\rho(\x)$, but rather a functional $\mathcal{P}$ which quantifies performance. To achieve this, a control update $\de\x^c$ is added to the passive update $\de\x^p$, so that their sum gives the total state update,
\begin{equation}
\de\x_t = \de\x^p_t + \de\x^c_t\left(\{\x\}_{-\infty}^t,\{\dxc\}_{-\infty}^{t-\de t}\right).
\end{equation}
Here $\dxc_t$ may depend on the entire history of the controlled system, $\{\x\}_{-\infty}^t$, and control updates $\{\dxc\}_{-\infty}^{t-\de t}$. 

Our goal is to bound the information required to achieve this control, which we quantify using the transfer entropy rate~\cite{schreiber2000measuring} from state $\x$ to control action $\de\x^c$:
\begin{equation}
\dot{I} \equiv \lim_{\de t\to0}\frac{I[\de\x^c_{t};\{\x\}_{-\infty}^t|\{\de\x^c\}_{-\infty}^{t-\de t}]}{\de t}.
\end{equation}
Here $I$ denotes conditional mutual information~\cite{Cover2006_Elements}. The transfer entropy rate quantifies the causal flow of information, accounting for memory via dependence on the past trajectories of both the system and control action.

Our first main result, derived in End Matter through information-theoretic accounting combined with the data-processing inequality, lower-bounds the information rate required for control. The bound is the passive entropy rate $\dot{S}_p$, the rate at which the passive dynamics increase the system Shannon entropy, $S[\x]\equiv\langle -\ln \rho(\x)\rangle_\rho$, immediately after control is turned off (Fig.~\ref{fig:schematic}):
\begin{equation}\label{eq:main_inequality}
\begin{aligned}
\dot{I}\geq \dot{S}_p & \equiv \lim_{\de t\to0}\frac{S\left[\x_t + \de\x^p_t\right] - S[\x_t]}{\de t}.
\end{aligned}
\end{equation}
Intuitively, the information required to control a system must at least match the rate at which the passive dynamics grow the volume in state space. This result is conceptually related to prior work~\cite{touchette2000information,touchette2004information}, but constitutes a significant advance: our bound relates the instantaneous passive entropy rate (rather than path-dependent entropies) to the control information rate $\dot{I}$, which quantifies causal information flow from system to controller. This result is powerful, since the passive entropy rate can generally be computed exactly for arbitrary $\rho(\x)$.

The bound~\eqref{eq:main_inequality} enables the derivation of performance-information Pareto frontiers, curves that quantify the best possible performance attainable with a given information rate. The performance metric of interest generally depends only on the steady-state distribution $\rho(\x)$, while the information rate $\dot{I}$ depends on the full history of the dynamics. This nonlocality makes it generally difficult to calculate, though recent progress has enabled tractable numerical computation~\cite{reinhardt2023path,das2025exact}. The passive entropy rate $\dot{S}_p$ depends only on $\rho(\x)$, and can thus more easily be related to the performance $\mathcal{P}$. 

The bound~\eqref{eq:main_inequality} is most useful when the objective is to tightly localize the steady-state probability distribution, as in the examples considered in this Letter. If instead the objective is to achieve a wider distribution than the passive dynamics would otherwise produce, $\dot{S}_p$ can be negative. We explore such a setup in SI Sec.~VI, showing that optimal control can be achieved with $\dot{I}=0$.

The problem of calculating the steady-state distribution $\rho(\x)$ that maximizes the performance $\mathcal{P}$ at fixed passive entropy rate $\dot{S}_p$ can be mapped onto the problem of finding the ground state of a corresponding Schrödinger equation (SI Sec.~IV). This mapping facilitates analytic or numerical calculations of both Pareto frontiers and optimal controlled steady-state distributions. 

To make our results fully general, we have omitted details of the controller and considered only the information flow from state $\x$ to control action $\de\x^c$. In some cases it is useful, however, to consider an explicit controller state $\bm{u}_t$. We show in SI Sec.~II that, so long as $\dxc_t$ depends only on $\bm{u}_t$, the passive entropy rate also lower-bounds the flow of information from system to the controller.

%%%%%%%%%%%%%%%%%%%%%%%%%%%%%%%%%%%%%%%%
\emph{Optimal Control Protocols.---}
Which protocols optimally control a system subject to a fixed information budget? To answer this question, we consider the gap in the bound~\eqref{eq:main_inequality} (SI Sec.~V), which reveals conditions that must be satisfied by optimal protocols. While Eq.~\eqref{eq:main_inequality} cannot always be saturated, we construct explicit optimal protocols for a broad class of systems. 

When the passive dynamics are state-independent, such that $p(\de\x^p_t|\x_t) = p(\de\x^p_t)$, the bound~\eqref{eq:main_inequality} can always be saturated, and optimal control is achieved by a protocol which probabilistically time-reverses the passive kernel $K_p(\x'|\x) = p(\x_t+\de\x^p_t|\x_t)$, with
\begin{equation}\label{eq:timereversalcontrol}
p(\de\x^c_t|\x_t) = \frac{K_p(\x_t|\x_t+ \de\x^c_t)p(\x_t+ \de\x^c_t)}{p(\x_{t})}.
\end{equation}
This is our second main theoretical result, providing an explicit optimal protocol for information-limited control when the passive dynamics are state-independent. Time-reversal protocols have been shown previously to minimize entropy production for thermodynamically consistent control of information engines~\cite{horowitz2011designing,horowitz2011thermodynamic}; here we show that probabilistic time reversal is the optimal protocol for a broad class of information-limited control problems regardless of thermodynamic considerations.

When the passive dynamics depend on the state, finding an optimal protocol is generally intractable outside of special cases such as linear dynamics (SI Sec.~VI). In the high-information limit when the steady-state distribution is highly localized, however, time-reversal protocols approach optimal performance since the passive dynamics are approximately state-independent in the narrow domain where steady-state probability is high. Combining this intuition with numerical exploration, we conjecture that the time-reversal protocol is near-optimal more generally. In SI Sec.~V we derive an explicit upper bound [Eq.~S50] on the gap between the information rate of the time-reversal protocol and the lower bound~\eqref{eq:main_inequality} given by the passive entropy rate.

\begin{figure}[t]
    \centering
    \includegraphics[width=\linewidth]{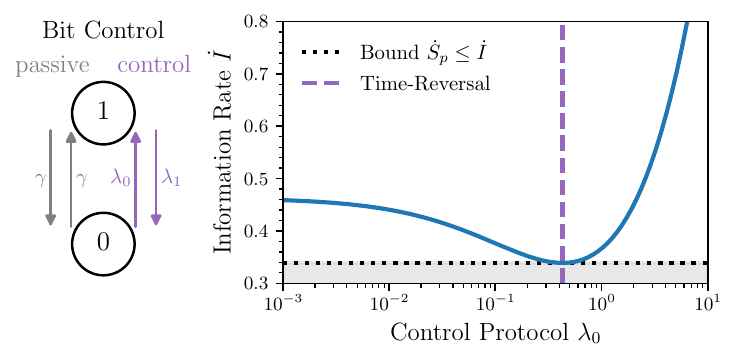}
    \caption{Controlling a noisy bit. Left: schematic, with passive bit flips at rate $\gamma$. The performance measure is $\mathcal{P} = \rho_0$, the steady-state probability of state 0. Right: information rate vs control protocol $\lambda_0$, always achieving a fixed performance $\rho_0=0.7$, with $\gamma=1$. Given $\rho_0$ and $\gamma$, $\lambda_0$ uniquely fixes $\lambda_1$. Black dotted line: Our bound $\dot{S}_p\leq\dot{I}$ [Eq.~\eqref{eq:main_inequality}]. Blue line: information rate $\dot{I}$. Purple dashed line: optimal rate $\lambda_0$ which saturates the bound by time-reversing the passive dynamics.
    \vspace{-3mm}
    }
    \label{fig:bitflip}
\end{figure}

%%%%%%%%%%%%%%%%%%%%%%%%%%%%
\emph{Controlling the State of a Bit.---}
For our first application, which provides intuition for the time-reversal control protocol, consider the problem of controlling a single bit (with state $\x\in\{0,1\}$) which passively flips its state at rate $\gamma$ [Fig.~\ref{fig:bitflip}]. Here the goal of the controller is to maximize the performance metric $\mathcal{P}=\rho_0$, the probability of the bit being in state $0$. 
%For mathematical convenience, we take the controller action to be addition by one modulo two. 
Evaluating Eq.~\eqref{eq:main_inequality}, the passive entropy rate gives a lower bound on the information rate required for control:
\begin{equation}
\dot{I} \geq \dot{S}_p = (2\rho_0-1)\gamma\ln\frac{\rho_0}{1-\rho_0}.
\end{equation}

Given $\gamma$ and a target performance $\rho_0$, the controller can tune the control jump rates $\lambda_0$ and $\lambda_1$ to minimize the information rate. With both $\gamma$ and $\rho_0$ fixed, the control protocol is fully characterized by a single parameter, which we take to be the rate $\lambda_0$ at which the controller induces bit errors. One might expect the optimal controller to make no errors ($\lambda_0=0$), acting only when necessary to flip the bit from the incorrect state $x=1$ at rate $\lambda_1$. Computing the information rate $\dot{I}$, however, reveals a minimum at a finite error rate $\lambda_0=\gamma(1-\rho_0)/\rho_0$ as predicted by Eq.~\eqref{eq:timereversalcontrol}. This is precisely the rate required to ensure that the control dynamics probabilistically time-reverse the passive dynamics, given the target $\rho_0$.

\begin{figure}[t]
    \centering
    \includegraphics[width=\linewidth]{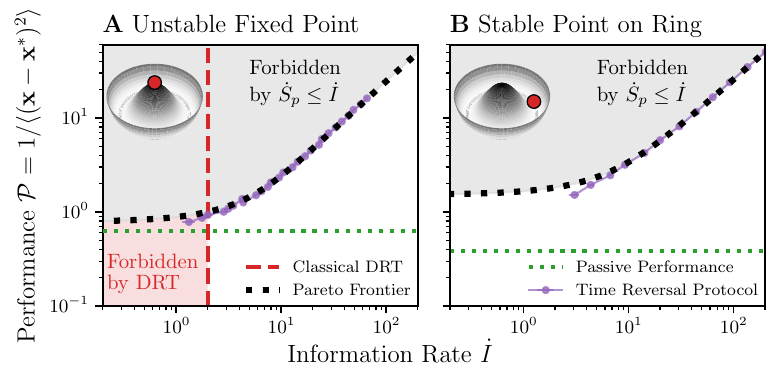}
    \caption{Controlling a Brownian particle in a Mexican--hat potential, with the target being either (a) the unstable fixed point $\x^*_1=(0,0)$, or (b) the stable fixed point $\x^*_2 = (1,0)$. With no control, the performance $\mathcal{P} = 1/\langle|\x-\x^*|^2\rangle$ remains finite (green dotted lines). The Data Rate Theorem (DRT) provides a lower bound on the information rate required for stabilization (red dashed line) only in the unstable case (a). Black dotted curves: Pareto frontier from Eq.~\eqref{eq:main_inequality}, with the forbidden region shaded in grey. Purple points: time-reversal protocol, Eq.~\eqref{eq:timereversalcontrol}, with information rate computed numerically using the TE-PWS algorithm~\cite{das2025exact}.
    \vspace{-3mm}
    }
    \label{fig:mexicanhat}
\end{figure}

%%%%%%%%%%%%%%%%%%%%%%%%%%%%
\emph{Localizing a Brownian Particle in a Potential.---}
A key advantage of our framework over past bounds such as the Data Rate Theorem is its applicability to the control of nonlinear and stochastic dynamical systems, which are ubiquitous in real-world control problems. Consider, as an example, the problem of localizing a Brownian particle in a nonlinear potential energy landscape, such as the two-dimensional Mexican--hat potential depicted in Fig.~\ref{fig:mexicanhat}. The passive dynamics comprise the drift imposed by the gradient of the potential, and diffusion. Suppose the goal is to tightly localize the system near some desired point $\x^*$, maximizing performance $\mathcal{P} = 1/\left\langle |\x-\x^*|^2\right\rangle_\rho$. We consider the control problem of minimizing the mean-squared distance from the fixed points $\x^*_1 = (0,0)$ (an unstable fixed point), and $\x^*_2 = (1,0)$ (a stable fixed point on the ring of minima).

We begin with the target $\x^*_1$, which is an unstable fixed point. Linearized around $\bm{x}_1^*$, both eigenvalues are unstable, so the classical Data Rate Theorem~\cite{tatikonda2004control} gives a minimum information rate required to achieve stable control. Our framework, by contrast, gives a full analytic performance-information Pareto frontier valid even below the classical threshold of stability [Fig.~\ref{fig:mexicanhat}(a)], as well as the distribution $\rho$ that maximizes ${\cal P}$ at fixed $\dot{S}_p$ (SI Sec.~VIII). Since the potential is confining for large $|\x|$, the performance remains finite under the passive dynamics, even with no control. Moreover, the time-reversal protocol is near-optimal for this system [Fig.~\ref{fig:mexicanhat}(a)].

For the second target $\x^*_2 = (1,0)$ [Fig.~\ref{fig:mexicanhat}(b)], which is stable under deterministic dynamics, the DRT provides no useful bound. Eq.~\eqref{eq:main_inequality}, by contrast, provides a full Pareto frontier for the problem which the time-reversal protocol saturates at high information. Thus our framework gives information limits for control even in cases where the DRT is completely uninformative.

\begin{figure}[t]
    \centering
    \includegraphics[width=\linewidth]{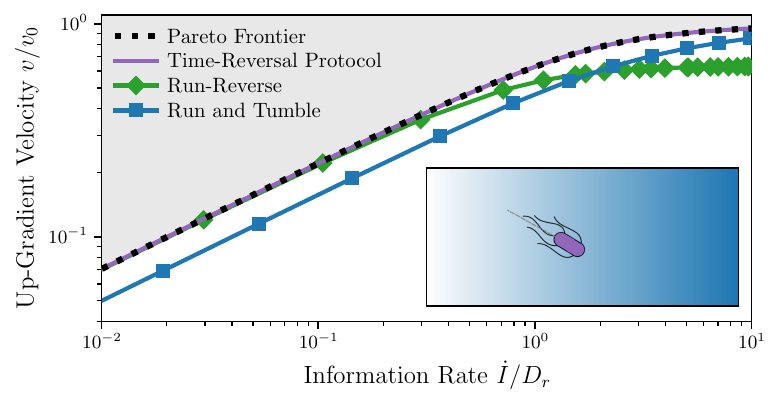}
    \caption{Up-gradient velocity vs information rate for navigation up a sensory gradient (inset: schematic depicting bacterial chemotaxis). Black dotted line: Pareto frontier, derived from Eq.~\eqref{eq:main_inequality} Purple: time-reversal protocol. Blue Squares: optimized Run-and-Tumble navigation. Green Diamonds: optimized Run-Reverse navigation, with information rate computed using TE-PWS~\cite{das2025exact}.
    \vspace{-3mm}
    }
    \label{fig:navigation}
\end{figure}

%%%%%%%%%%%%%%%%%%%%%%%%%%%%%%%%%%%%%%%%%%%%%%
\emph{Navigation up a Sensory Gradient.---}
Bacteria and other microbes must navigate their environments with limited information gathered through noisy sensors~\cite{mattingly2021escherichia}, raising the question of how quickly they can navigate up a gradient with a given amount of information. Following recent work~\cite{betancourt2026discrete}, we consider an agent navigating up a sensory gradient in two dimensions. The agent moves at a constant speed $v_0$, but can sense and control its heading angle $\theta\in[0,2\pi)$ with respect to the gradient direction. The angle $\theta$ suffers passive rotational diffusion with coefficient $D_r$. The agent seeks to maximize performance quantified by the up-gradient velocity, $\mathcal{P} = v/v_0=\langle\cos\theta\rangle_\rho$. Our bound provides an exact velocity--information Pareto frontier (SI Sec.~IX) [Fig.~\ref{fig:navigation}], generalizing past work~\cite{betancourt2026discrete} to allow arbitrary control protocols with memory. 

The bound is saturated by the probabilistic time-reversal protocol~\eqref{eq:timereversalcontrol}, which applies a nonlinear steering force (SI Sec.~IX) to maximize velocity at a given information rate without using memory of past states or actions. This optimal protocol, however, is beyond the capabilities of the bacteria that motivated the setup, which typically measure the time derivative of the concentration (or its logarithm), analogous to measuring $\cos\theta$ in our simplified model. Given this constraint, a common navigation strategy in the microbial world (employed by, e.g., \emph{E. coli}~\cite{mattingly2021escherichia}) is to Run-and-Tumble, which performs markedly worse than the optimal protocol [Fig.~\ref{fig:navigation}]. 

Given the observability constraint of measuring only $\cos\theta$, and allowing arbitrary control actions and memory, we prove in SI Sec.~IX that in the low-information limit, the Pareto frontier is asymptotically saturated by an instantaneous Run-Reverse strategy that does not use memory [Fig.~\ref{fig:navigation}]. This navigation strategy, implemented by engaging in turns of $\Delta\theta=\pi$ at a heading-dependent rate $\lambda(\theta)$, is employed by real-world bacteria such as \emph{P. aeruginosa} and \emph{V. cholerae}~\cite{son2016speeddependent}. Given that \emph{in vivo} information rates for bacterial chemotaxis have been estimated at $\mathcal{O}(10^{-2})$ bit/s~\cite{mattingly2021escherichia}, it is thus plausible that these microbes have evolved to navigate using the optimal control strategy given their limited sensory information and constrained observability. Beyond the low-information limit, memory may yet improve performance.

%%%%%%%%%%%%%%%%%%%%%%%%%%%%%%%%%%%%%%%%%%%%%%
\emph{Extracting Energy using Information.---}
As a final application, we consider the information cost of operating information engines, which lie at the heart of the intersection between information thermodynamics and control theory~\cite{paneru2020colloidal,du2024performance}. These devices consist of a controller which uses information to feedback-control a physical system to extract heat energy (at a rate $\dot{Q}$) from a thermal bath without directly performing work on the system. The second law of thermodynamics imposes a lower-bound on the information rate required for control, but can be extremely loose~\cite{leighton2024information}. Our framework provides a path to interrogate the fundamental exchange rate between information and energy.

To explore such information costs, consider the information engine realized experimentally in Ref.~\cite{saha2021maximizing}. The controlled system is a Brownian particle with scaled mass $\delta_g$ in a harmonic laser trap with relaxation timescale $\tau_r$. The controller updates the position of the trap to raise the particle up against the downward force of gravity, while the particle passively relaxes within the trap. To realize an information engine, the controller must on average do no work on the particle, i.e., the controller cannot on average change the trap potential energy. Performance is quantified by the rate, $\mathcal{P} = \beta\dot{Q}$, at which heat is extracted from the bath, equal by the first law of thermodynamics to the rate of free energy storage as gravitational potential energy. We take the state of the system to be the difference between the particle position and trap center, which reaches a true steady state.

The second law of information thermodynamics requires a minimum information rate $\dot{I}\geq\beta\dot{Q}$, leaving open the possibility that the maximum rate of heat extraction could be achieved using only a finite information rate. Applying Eq.~\eqref{eq:main_inequality}, we derive a tighter Pareto frontier bounding the rate at which free energy can be extracted for a given information rate (SI Sec.~X):
\begin{equation}\label{eq:info_engine_frontier}
\beta\dot{Q}  + \tau_r\left(\frac{\beta\dot{Q}}{\delta_g}\right)^2\leq \frac{\dot{I}}{1+ \tau_r\dot{I}}.
\end{equation}
This bound reduces to the second law in the low-information (and thus low-$\dot{Q}$) limit, but the right-hand side asymptotically approaches $\tau_r^{-1}$ as $\dot{I}\to\infty$, such that the theoretical maximum performance requires a diverging information rate.

\begin{figure}[t]
    \centering
    \includegraphics[width=\linewidth]{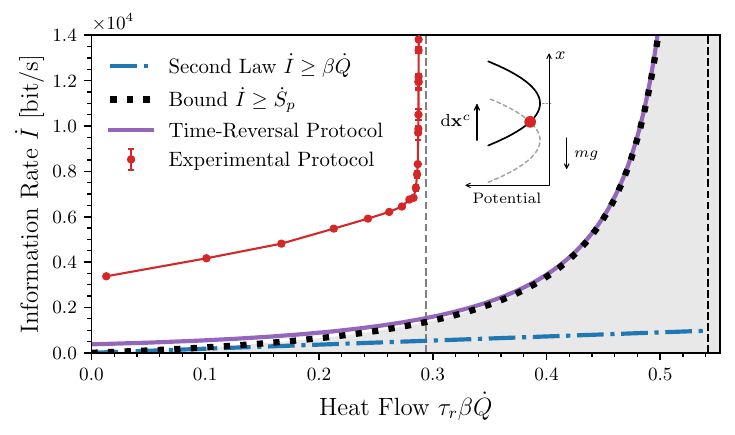}
    \caption{Information rate vs heat extracted from the bath for an information engine. 
    Inset: a massive Brownian bead in a feedback-controlled vertical laser trap.
    Blue: second law of information thermodynamics. Black dotted line: Pareto frontier~\eqref{eq:info_engine_frontier}. Red points: simulations of the experimentally implemented control protocol from Ref.~\cite{saha2021maximizing} plotted parametrically, varying the measurement noise $\sigma_m$, with information rate evaluated using TE-PWS~\cite{das2025exact}. Purple line: Time-reversal protocol. Vertical dashed lines: maximum heat flow with no average work (black) and with no work for each controller action (grey). Parameters are $t_s/\tau_r= 1/41$, $\delta_g=0.8$, and $\tau_r = 0.8\times 10^{-3}$s, from the experimental setup in Ref.~\cite{saha2022bayesian}.
    \vspace{-3mm}
    }
    \label{fig:infoengine}
\end{figure}

%In the experimental implementation of Ref.~\cite{saha2021maximizing}, the controller measures the particle position $\hat{x}_t = x_t + \mathcal{N}(0,\sigma_m^2)$ (with measurement noise $\sigma_m$) at periodic time intervals of length $t_s$, and moves the trap center $z$ by a distance $\de z^c$ which depends on the controller's estimate of $u_t$. The control protocol updates the trap center based only on the measured particle position as $\de z^c_{t = nt_s} = \alpha(\hat x_t-z_t)$ if $\hat{x}_t>z_t$, and $\de z^c_{t = nt_s}=0$ otherwise, for integer $n$, and feedback gain parameter $\alpha$ that is tuned to ensure the controller does no average work. For simplicity, we omit the feedback delay in Ref.~\cite{saha2021maximizing}; this would be a useful direction for future work.

The prediction of diverging information cost is confirmed in simulations for the control protocol implemented experimentally in Ref.~\cite{saha2021maximizing} (Fig.~\ref{fig:infoengine}). Because the protocol is restricted such that the only possible trap center movement is a reflection about the particle position, the maximum free energy extraction is lower than predicted by Eq.~\eqref{eq:info_engine_frontier}. Relaxing this restriction (allowing temporary energy exchange between the particle and the controller~\cite{lucero2021maximal}) unlocks significant performance improvements. Probabilistically time-reversing the passive dynamics (SI Sec.~X) yields a near-optimal protocol which dramatically outperforms the experimentally implemented protocol (Fig.~\ref{fig:infoengine}).

\newpage
%%%%%%%%%%%%%%%%%%%%%%%%%%%%%%
\emph{Discussion.---}
In this \type we showed that the information rate required for feedback control of stochastic systems is lower-bounded by the passive entropy rate, which depends only on the uncontrolled passive dynamics and the target steady-state distribution $\rho(\x)$. The passive entropy rate can easily be related to performance metrics of interest, making Eq.~\eqref{eq:main_inequality} a powerful tool for deriving information--performance Pareto frontiers for varied control problems. 

Examining saturation conditions for the bound~\eqref{eq:main_inequality} leads to an explicit optimal control protocol (Eq.~\eqref{eq:timereversalcontrol}) for state-independent passive dynamics: probabilistic time-reversal. More generally, time-reversal protocols are near-optimal for many applications. Since optimal protocols are inherently probabilistic, they can in many cases be implemented even in the presence of noise arising from measurement, decision-making, or actuation. For real-world implementation, probabilistic protocols may be costly to implement on deterministic digital computers, but are ideally suited for devices such as thermodynamic computers~\cite{melanson2025thermodynamic,jelinvcivc2025efficient} or physical learning systems~\cite{momeni2025training,stern2025physical}. 

Our results constitute a significant advance from the Data Rate Theorem~\cite{tatikonda2004control,nair2004topological,nair2007feedback}, generalizing to arbitrary stochastic dynamics, producing full Pareto frontiers rather than just thresholds for stabilization, and providing finite bounds even when the DRT suggests no information is required for control. In applications, our framework gives principled explanations for the behavior of chemotactic microbes, and shows that the information cost to operate a feedback-control information engine diverges where the second law requires only a finite cost.

Here we focused on feedback control to maintain target steady states. Time-dependent control constitutes another important class of problems in control theory, where the goal of a control protocol is to move the system from one state to another in finite time. Minimizing the first-passage time is also an important distinct objective relevant in many areas of biology. Generalizing our framework to such time-dependent control problems will be an important direction for future work. Given the prevalence of long-range non-Markovian memory throughout biology~\cite{leighton2025decomposing}, exploring the control of non-Markovian dynamical systems could also prove insightful.

This work focused on the information cost of control, ignoring the energetic cost. Our approach, being agnostic to thermodynamic considerations, is well suited for application to complex biological systems where the coupling between control and energetic costs is not easily identified. Nonetheless, energetic constraints are central to many important control problems, so understanding the relationship between information and energetic costs of control is critical. When the control update depends only on a resolved controller state, we show in SI Sec.~II that the passive entropy rate bounds not only the information rate, but also the energetic cost of the controller.

Finally, this \type necessarily considered only a handful of illustrative applications. We envision the present framework and bound for information-limited control being applicable to control problems across biology, chemistry, physics, engineering, and economics, and hope to see such applications explored in future work.

\emph{Acknowledgements}.---We thank Kevin Chen, David Hathcock, Bert Kappen, Artemy Kolchinsky, Chris Lynn, Henry Mattingly, Eric Regis, Jordan Sawchuk, Sekhar Tatikonda, and Nick Weaver for helpful discussions and/or feedback on the manuscript. We acknowledge the use of AI tools to assist with implementing some of the numerical calculations, in particular adapting the TE-PWS algorithm from Ref.~\cite{das2025exact} for our use case.

\emph{Funding}.---This work was supported by Mossman and NSERC Postdoctoral Fellowships (M.P.L.), the Yale Program in Physical and Engineering Biology (J.M.B.), NIH grants R35GM158058 (T.E., M.C.A.) and R35GM138341 (B.B.M., M.C.A.), and an Alfred P. Sloan Foundation \emph{Matter-to-Life} grant (T.E., B.B.M., M.C.A.).

\emph{Code Availability}.---Python code for reproducing all figures in the manuscript is available on GitHub~\cite{github}.

%%%%%%%%%%%%%%%%%%%%%%%%%%%%%%%%%%%%%%%%%%%%%%%%
\bibliography{main}% Produces the bibliography via BibTeX.

\section*{End Matter}
Here we derive our bound, Eq.~\eqref{eq:main_inequality}, on the information rate required for feedback control. As in the main text, consider a stochastic system with state $\x$. In the absence of control, it follows Markovian dynamics that we refer to as ``passive", indicated by superscript $p$, such that the passive dynamics produce an update $\de\x^p_t = \x_{t+\de t}-\x_t$ over time $\de t$ distributed according to the conditional distribution $p(\de \x^p_t|\x_t)$. As an explicit example, the passive dynamics could be described by the following stochastic differential equation:
\begin{equation}\label{General_SDE}
\de\x^p_t = \underbrace{\bm{F}(\x)\de t}_\mathrm{drift} + \underbrace{\sqrt{2\diffusivity}\,\de\bm{W}^p_t}_\mathrm{diffusion} + \underbrace{\Delta\x\, \mathrm{d}J(\lambda(\x,\Delta\x))}_\mathrm{jumps}.
\end{equation}
Here $\bm{F}(\x)$ is a drift force, $\diffusivity$ is a diffusion tensor, and $\bm{W}^p_t$ denotes a standard Wiener process. The last term represents discrete jumps of any distance $\Delta \x$, governed by a Poisson process $\mathrm{d}J$ with rate $\lambda(\x,\Delta\x)$. We assume the state space has no reflecting boundaries. These dynamics~\eqref{General_SDE} can equivalently be described by a Fokker--Planck equation for the time-dependent probability $p_t(\x)$,
\begin{equation}\label{General_FPE}
\begin{aligned}
& \frac{\partial}{\partial t} p_t(\x) = -\nabla\!\cdot\! \left[\bm{F}(\x) p_t(\x)\right] - \! \int\!\de\Delta\x\, \lambda(\x,\Delta\x)p_t(\x)\\
& + \int\!\de\Delta\x\, \lambda(\x\!-\!\Delta\x,\Delta\x)p_t(\x\!-\!\Delta\x) +\nabla\cdot\diffusivity\nabla p_t(\x).
\end{aligned}
\end{equation}

We seek to control this system to maintain a steady-state distribution $\rho(\x)$ by adding a control update $\de\x^c$ to the passive update $\de\x^p$, so that their sum gives the total state update,
\begin{equation}
\de\x_t = \de\x^p_t + \de\x^c_t\left(\{\x\}_{-\infty}^t,\{\dxc\}_{-\infty}^{t-\de t}\right).
\end{equation}
Here $\dxc_t$ may depend on the entire history of the controlled system, $\{\x\}_{-\infty}^t$, and control updates $\{\dxc\}_{-\infty}^{t-\de t}$. 

To derive Eq.~\ref{eq:main_inequality}, consider the control and passive updates of the state applied in sequence:
\begin{equation}\label{eq:arrowdiagram}
\x_t \underset{+ \dxc_t}{\longrightarrow} \y_t \underset{+\de\x^p_t}{\longrightarrow} \x_{t+\de t}
\end{equation}
Here we have chosen an ordering of the two updates, but the result can also be derived with the opposite ordering, and in any case the ordering does not matter in the limit as $\de t\to0$. The intermediate state $\y_t \equiv \x_t + \dxc_t$ may depend on the full history of the system $\xtraj$ as well as the control history $\dxctraj$.
The passive update $\de\x^p_t$, by contrast, depends only on $\y_t$ since the passive dynamics are Markovian. 

Now consider the increase in entropy due to the passive dynamics, $\Delta S_p \equiv S[\x_{t+\de t}] - S[\y_t]$. Assuming stationarity, this must equal the entropy decrease due to active control, $\Delta S_p = S[\x_t] - S[\y_t]$. Note next that the marginal entropies of $\x_t$ and $\y_t$ can be expanded as
\begin{equation}\label{eq:Sx_expansion}
S[\x_t] = S[\x_t|\{\dxc\}_{-\infty}^{t-\de t}] + I[\x_t;\{\dxc\}_{-\infty}^{t-\de t}],
\end{equation}
and 
\begin{equation}\label{eq:Sy_expansion}
\begin{aligned}
S[\y_t] & = S[\y_t|\{\dxc\}_{-\infty}^t] + I[\y_t;\{\dxc\}_{-\infty}^t] \\
& = S[\x_t|\{\dxc\}_{-\infty}^t] + I[\y_t;\{\dxc\}_{-\infty}^t].
\end{aligned}
\end{equation}
In the second line, we used the fact that $\y_t$ is a deterministic translation of $\x_t$ given $\dxc_t$. We then expand the passive entropy change using Eqs.~\eqref{eq:Sx_expansion} and \eqref{eq:Sy_expansion} to obtain
\begin{equation}
\begin{aligned}
\Delta S_p & = S[\x_{t}] - S[\y_t]\\
& = I[\dxc_t;\x_t|\{\dxc\}_{-\infty}^{t-\de t}] \\
 &\maybespace + I[\x_t;\{\dxc\}_{-\infty}^{t-\de t}] - I[\y_t;\{\dxc\}_{-\infty}^t].
\end{aligned}
\end{equation}
In the second line, we identified the conditional mutual information $I[\dxc_t;\x_t|\{\dxc\}_{-\infty}^{t-\de t}]$ as the difference in conditional entropies $S[\x_t|\{\dxc\}_{-\infty}^{t-\de t}] - S[\x_t|\{\dxc\}_{-\infty}^t]$. Next, we note that the data-processing inequality~\cite{Cover2006_Elements} implies $I[\y_t;\{\dxc\}_{-\infty}^t]\geq I[\x_{t+\de t};\{\dxc\}_{-\infty}^t]$. We also note that $ I[\dxc_t;\x_t|\{\dxc\}_{-\infty}^{t-\de t}]\leq I[\dxc_t;\{\x\}_{-\infty}^t|\{\dxc\}_{-\infty}^{t-\de t}]$ since $\{\x\}_{-\infty}^t$ cannot be less informative about $\dxc_t$ than $\x_t$ alone. Together, these give
\begin{equation}
\begin{aligned}
\Delta S_p & \leq I[\dxc_t;\{\x\}_{-\infty}^t|\{\dxc\}_{-\infty}^{t-\de t}]\\
&\maybespace + I[\x_t;\{\dxc\}_{-\infty}^{t-\de t}] - I[\x_{t+\de t};\{\dxc\}_{-\infty}^t].
\end{aligned}
\end{equation}
Since the system is at steady state, the terms in the second line are equal so that their difference is zero, leaving
\begin{equation}
\Delta S_p \leq I[\dxc_t;\{\x\}_{-\infty}^t|\{\dxc\}_{-\infty}^{t-\de t}].
\end{equation}
Dividing by $\de t$ and taking the limit $\de t\to 0$ leads to Eq.~\ref{eq:main_inequality}.

\end{document}

\title{Supplementary Information for ``Passive Entropy Rate Bounds the Information Required for Feedback Control"}

\author{Matthew P. Leighton}
\email{matthew.leighton@yale.edu}
\affiliation{\physics}
\author{Jose M. Betancourt} 
\affiliation{\physics}
\author{Thierry Emonet} 
\affiliation{\physics}
\affiliation{\mcdb}
\author{Benjamin B. Machta}
\affiliation{\physics}
\author{Michael C. Abbott}
\affiliation{\physics}

\date{\today}

\maketitle
\tableofcontents

\newpage

\section{A Note on Notation}
In the main text we quantify the information used by the controller using the transfer entropy rate~\cite{schreiber2000measuring} from state $\x$ to control action $\de\x^c$:
\begin{equation}
\dot{I} \equiv \lim_{\de t\to0}\frac{I[\de\x^c_{t};\{\x\}_{-\infty}^t|\{\de\x^c\}_{-\infty}^{t-\de t}]}{\de t}\equiv \dot{\mathcal{T}}_{\x\to\de\x^c}.
\end{equation}
While in the main text we denoted this transfer entropy rate $\dot{I}$ for notational simplicity, here and throughout the SI we use the notation $\dot{\mathcal{T}}_{\x\to\de\x^c}$ in order to distinguish between different information rates under consideration.

\section{Considering an Explicit Controller State}\label{app:explicitcontroller}
Here we consider information-limited control in the case where an explicit controller state and dynamics are resolved. Consider a controlled system with state $\x$, and a controller with state $\bu$. Let us assume their joint dynamics are bipartite and Markovian, and consider the steady state. We make no further assumptions regarding the dynamics of the controller state, but write the dynamics of the controlled system as
\begin{equation}
\de\x_t = \de\x_t^p(\x_t) + \de\x_t^c(\x_t,\bu_t).
\end{equation}
Here $\de\x_t^p$ is the passive update, which depends only on the current state $\x_t$, and $\de\x_t^c$ is the control update, which may in general depend on both the system state $\x_t$ and controller state $\bu_t$. We are interested in deriving bounds on two different information rates. The first is the information flow $\dot{I}_{\bu}$~\cite{horowitz2014thermodynamics,ehrich2023energy}, which quantifies the rate at which the controller dynamics change the mutual information between the controller and the controlled system:
\begin{equation}
\dot{I}_{\bu} \equiv \lim_{\de t\to0}\frac{I[\x_t;\bu_{t+\de t}] - I[\x_t;\bu_{t}]}{\de t}.
\end{equation}
The second is the transfer entropy rate from $\x$ to $\bu$,
\begin{equation}
\dot{\mathcal{T}}_{\x\to\bu} \equiv \lim_{\de t\to0}\frac{1}{\de t}I\left[\bu_{t+\de t};\{\x\}_{-\infty}^t|\{\bu\}_{-\infty}^t\right].
\end{equation}
Given the assumption of bipartite Markovian dynamics, these are related by $\dot{I}_{\bu}\leq\dot{\mathcal{T}}_{\x\to\bu}$~\cite{hartich2017stochastic}. We will now connect these information rates to the passive entropy rate $\dot{S}_p$.

First consider mutual information balance at steady state. The rate $\dot{I}_{\bu}$ at which the controller dynamics increase the mutual information must equal the rate $-\dot{I}_{\x}$ at which the controlled system dynamics decrease it. We can in turn decompose this rate following standard methods in bipartite stochastic thermodynamics~\cite{horowitz2014thermodynamics,leighton2024information}:
\begin{align}
\dot{I}_{\bu} & = -\dot{I}_{\x}\\
& = -\dot{I}_{\x}^p - \dot{I}_{\x}^c\\
& = -\dot{I}_{\x}^p - \dot{S}_{\x}^c + \dot{S}_{\x|\bu}^c.
\end{align}
Here $\dot{I}_{\x}^p$ and $\dot{I}_{\x}^c$ are the rates at which the passive and controlled updates of $\x$ respectively change the mutual information between $\x$ and $\bu$, $\dot{S}_{\x}^c$ is the rate at which the controlled update changes the marginal entropy of $\x$, and $\dot{S}_{\x|\bu}^c$ is the rate at which the controlled update changes the conditional entropy of $\x$ given $\bu$. Since the marginal entropy of $\x$ is constant, we must have $\dot{S}_{\x}^c + \dot{S}_{p}=0$. Thus we have
\begin{equation}
\dot{S}_{p} = \dot{I}_{\bu} + \dot{I}_{\x}^p - \dot{S}_{\x|\bu}^c.
\end{equation}
Since the passive dynamics are independent of the controller state $\bu$, they can only decrease the mutual information, so that $\dot{I}_{\x}^p\leq0$. Further, the conditional entropy rate can be bounded by
\begin{align}
\dot{S}_{\x|\bu}^c & = \lim_{\de t\to0}\frac{S[\x_t + \de\x_t^c|\bu_t] - S[\x_t|\bu_t]}{\de t}\\
& = \lim_{\de t\to0}\frac{S[\x_t + \de\x_t^c|\bu_t] - S[\x_t|\bu_t]}{\de t}\\
& - \lim_{\de t\to 0}\frac{S[\x_t + \de\x^c_t|\de\x_t^c,\bu_t] - S[\x_t|\de\x_t^c,\bu_t]}{\de t}\\
& = \lim_{\de t\to0}\frac{I[\x_t + \de\x_t^c;\de\x_t^c|\bu_t] - I[\x_t;\de\x_t^c|\bu_t]}{\de t}\\
& \geq - \lim_{\de t\to0}\frac{I[\x_t;\de\x_t^c|\bu_t]}{\de t}.
\end{align}
Thus we have
\begin{equation}
\dot{S}_p\leq \dot{I}_{\bu} + \lim_{\de t\to0}\frac{I[\x_t;\de\x_t^c|\bu_t]}{\de t}.
\end{equation}
From this inequality we see that when the control update depends only on the controller state, such that $\de\x^c_t(\x_t,\bu_t) = \de\x_t^c(\bu_t)$, we have
\begin{equation}
\dot{S}_p\leq \dot{I}_{\bu} \leq \dot{\mathcal{T}}_{\x\to\bu}.
\end{equation}
Here the second inequality follows from Ref.~\cite{hartich2017stochastic}. Under this same condition on the control update, data processing~\cite{derpich2021directed} gives $\dot{\mathcal{T}}_{\x\to\de \x^c}\leq \dot{\mathcal{T}}_{\x\to\bu}$, so that we have the following hierarchy of control information rates:
\begin{equation}\label{eq:inforatehierarchy}
\dot{S}_p\leq \min\left\{\dot{I}_{\bu},\dot{\mathcal{T}}_{\x\to\de \x^c}\right\}\leq\max\left\{\dot{I}_{\bu},\dot{\mathcal{T}}_{\x\to\de \x^c}\right\}\leq \dot{\mathcal{T}}_{\x\to\bu}.
\end{equation}
There is not a general ordering between $\dot{I}_{\bu}$ and $\dot{\mathcal{T}}_{\x\to\de \x^c}$, and further these orderings are not guaranteed if $\de\x^c_t$ depends directly on $\x_t$ in addition to its dependence on $\bm{u}$ (the main result of the main text, Eq.~\eqref{eq:main_inequality}, still holds in this case).

The result Eq.~\eqref{eq:inforatehierarchy} has an important consequence for the thermodynamic cost of operating the controller, assuming the controller operates in contact with a thermal bath at temperature $T$. Since the joint system-controller dynamics are bipartite and Markovian, the heat dissipation rate of the controller $\dot{Q}_{\bu}$ is lower-bounded by the information flow $\dot{I}_{\bu}$~\cite{horowitz2014second,ehrich2023energy} such that (so long as $\de\x^c_t$ depends only on $\bu_t$)
\begin{equation}
\dot{S}_p \leq \frac{1}{k_BT}\dot{Q}_{\bu}.
\end{equation}
Thus in the case where the control update depends only on the controller state, our Pareto frontiers from the main text also bound the relationship between performance and the energetic cost of feedback control.

\section{Computing and Bounding Information Rates}\label{app:inforatecomputation}
Here we provide general details for computing and bounding the control transfer entropy rate, defined as
\begin{equation}
\dot{\mathcal{T}}_{\x\to\de\x^c} \equiv \lim_{\de t\to0}\frac{I[\de\x^c_{t};\{\x\}_{-\infty}^t|\{\de\x^c\}_{-\infty}^{t-\de t}]}{\de t}.
\end{equation}
Evaluating the conditional mutual information gives
\begin{equation}\label{eq:SITEdef}
\dot{\mathcal{T}}_{\x\to\de\x^c} = \lim_{\de t\to0}\frac{1}{\de t} \left\langle\left\langle\ln\frac{p\left(\de\x^c_{t}|\{\x\}_{-\infty}^t,\{\de\x^c\}_{-\infty}^{t-\de t}\right)}{p\left(\de\x^c_{t}|\{\de\x^c\}_{-\infty}^{t-\de t}\right)}  \right\rangle_{p\left(\de\x^c_{t}|\{\x\}_{-\infty}^t,\{\de\x^c\}_{-\infty}^{t-\de t}\right)} \right\rangle_{p\left(\{\x\}_{-\infty}^t,\{\de\x^c\}_{-\infty}^{t-\de t}\right)}.
\end{equation}
Thus evaluating the information rate requires knowledge of the control update distribution given the full history of both the system and controller, $p\left(\de\x^c_{t}|\{\x\}_{-\infty}^t,\{\de\x^c\}_{-\infty}^{t-\de t}\right)$, as well as the distribution of the control update given only the control history, $p\left(\de\x^c_{t}|\{\de\x^c\}_{-\infty}^{t-\de t}\right)$. This second distribution is generally extremely difficult to compute outside of special cases.

To address the difficulty of computing the full-history transfer entropy rate, we show that the instantaneous control information rate,
\begin{equation}\label{eq:localcontrolinfo}
\dot{I}^+_c = \lim_{\de t\to 0}\frac{I[\x_t;\de\x^c_t]}{\de t},
\end{equation}
is an upper bound on $\dot{\mathcal{T}}_{\x\to\de\x^c}$ for control dynamics which do not rely on memory. Without memory the control is Markovian, with $p(\dxc_t|\{\x\}_{-\infty}^t,\{\dxc\}_{-\infty}^{t-\de t}) = p(\dxc_t|\x_t)$. In this case the transfer entropy rate obeys
\begin{equation}\label{eq:controlinfoupperbound}
\begin{aligned}
\dot{\mathcal{T}}_{\x\to\de\x^c} & = \lim_{\de t\to0}\frac{I[\de\x^c_t;\{\x\}_{-\infty}^t|\{\de\x^c\}_{-\infty}^{t-\de t}]}{\de t}\\
& = \lim_{\de t\to0}\frac{I[\de\x^c_t;\x_t|\{\de\x^c\}_{-\infty}^{t-\de t}]}{\de t}\\
& = \lim_{\de t\to 0}\frac{1}{\de t}\left( I[\de\x^c_t;\x_t] - I[\de\x^c_t;\{\de\x^c\}_{-\infty}^{t-\de t}] + I[\de\x^c_t;\{\de\x^c\}_{-\infty}^{t-\de t}|\x_t]\right)\\
& = \dot{I}^+_c - \lim_{\de t\to0}\frac{I[\de\x^c_t;\{\de\x^c\}_{-\infty}^{t-\de t}]}{\de t} + 0\\
&\leq \dot{I}^+_c.
\end{aligned}
\end{equation}
This also illustrates that the instantaneous control information and transfer entropy rate are equal when $I[\de\x^c_t;\{\de\x^c\}_{-\infty}^{t-\de t}]=0$, i.e., the control history alone contains no information about the next control update.

If instead the control strategy is non-Markovian, then this inequality need not hold. If we expand the state space to include a memory variable $M_t$ such that the control depends only on $\x_t$ and $M_t$, then the preceding argument implies  
\begin{equation}
\dot{\mathcal{T}}_{\x\to\de\x^c}\leq \lim_{\de t\to 0}\frac{I[\x_t,M_t;\de\x^c_t]}{\de t}.
\end{equation}

The instantaneous control information rate~\eqref{eq:localcontrolinfo} is particularly useful because the information rate $\dot{I}_c^+$ is generally straightforward to compute. Suppose the control takes the form
\begin{equation}
\dxc_t = \bm{F}_c(\x_t)\de t + \sqrt{2\diffusivity_c}\,\de\bm{W}_t^c + \Delta\x\, \de J\left(\lambda_c(\x,\Delta\x)\right),
\end{equation}
which can accommodate the optimal probabilistic time-reversal protocol. Then the instantaneous control information rate can be evaluated as~\cite{betancourt2026discrete}
\begin{equation}\label{eq:localcontrolinfoevaluated}
\dot{I}^+_c = \frac{1}{4}\left[\left\langle\bm{F}_c\cdot \diffusivity_c^{-1}\bm{F}_c\right\rangle_\rho - \left\langle\bm{F}_c\right\rangle_\rho\cdot\diffusivity_c^{-1}\left\langle \bm{F}_c\right\rangle_\rho\right] + \iint\de\x\de\Delta\x\, \rho(\x)\lambda_c(\x,\Delta\x)\ln\frac{\lambda_c(\x,\Delta\x)}{\bar{\lambda}_c(\Delta\x)},
\end{equation}
where $\bar{\lambda}_c(\Delta\x) = \int\de\x \,\rho(\x)\lambda_c(\x,\Delta\x)$.

%%%%%%%%%%%%%%%%%%%%%%%%%%%%%%%%%%%%%%%%%%%%%%%%%%%%%%%%%%%%%%%%%
\section{Deriving Pareto Frontiers}\label{app:schrodinger_mapping}
Here we develop a general procedure for deriving performance-information Pareto frontiers for information-limited control problems. We begin with the passive entropy rate, which for the passive dynamics~\eqref{General_SDE} we write as an integral over $\x$:
\begin{equation}\label{eq:explicitpassiveentropy}
\dot{S}_p = \int\de\x \,\rho(\x)\left[(\nabla\ln \rho(\x))\cdot\bm{D}\,\nabla\ln \rho(\x) + \nabla\cdot \bm{F}(\bm{x})  + \int\de\Delta\x\, \lambda(\x,\Delta\x)\ln\frac{\rho(\x)}{\rho(\x+\Delta\x)} \right].
\end{equation}
We further assume the performance metric can be written as the average of some scalar function $A(\x)$:
\begin{equation}
\mathcal{P}[\rho] = \int\de\x \rho(\x) A(\x).
\end{equation}
Since we have shown the bound~\eqref{eq:main_inequality} between passive entropy rate and information rate can be tight, the Pareto frontier for a given problem can be determined by extremizing $\mathcal{P}$ at fixed $\dot{S}_p$. This can be achieved using the objective function
\begin{equation}
\mathcal{L} = \mathcal{P}[\rho] - \gamma\dot{S}_p + \alpha\int\de\x \rho(\x).
\end{equation}
Here $\gamma$ and $\alpha$ are Lagrange multipliers, respectively enforcing constant $\dot{S}_p$ and normalization of $\rho(\x)$. The condition for extremizing $\mathcal{L}$ is derived by setting the functional derivative $\delta\mathcal{L}/\delta\rho=0$. 

In the absence of passive jumps, this is a standard variational calculus problem. Defining $\psi(\x)$ such that $\psi^2(\x)=\rho(x)$, $\lambda\equiv 1/\gamma$, and $E_0 = \alpha/\gamma$, the Euler--Lagrange equation gives 
\begin{equation}
\left[-4\nabla\cdot\diffusivity\nabla + \nabla\cdot\bm{F}(\x) -\lambda A(\x)\right]\psi(\x) = E_0(\lambda)\psi(x).
\end{equation}
This can be mapped onto a Schrödinger equation by defining the potential $U_\lambda(\x) = \nabla\cdot\bm{F}(\x) - \lambda A(\x)$. If the diffusion matrix is diagonal and isotropic ($\diffusivity=D\bm{I}$), then we have
\begin{equation}\label{eq:schrodinger}
\underbrace{\left[-4D\nabla^2 + U_\lambda(\x)\right]}_{\equiv H_\lambda}\psi(\x) = E_0(\lambda)\psi(\x),
\end{equation}
where we have defined the Hamiltonian $H_\lambda$. This is a linear, time-independent Schrödinger equation, and suggests identification of the normalization Lagrange multiplier $E_0$ with the ground-state energy of the Hamiltonian $H_\lambda$, while the square of the ground-state wavefunction $\psi_0(\x)$ gives the probability distribution $\rho(\x)$ which extremizes $\mathcal{L}$. The ground state is the solution which extremizes the objective function.

The mapping can still be performed if $\diffusivity$ is non-diagonal and/or anisotropic through a change of variables. Since the diffusion matrix $\diffusivity$ must be positive semidefinite, it can always be scaled out by changing to a variable $\y$ satisfying $\x = [(\diffusivity + \diffusivity^\top)/2]^{1/2}\y$. This would then yield a linear and time-independent Schrödinger equation in the variable $\y$.  Finally, in the case where passive jumps are present, an additional term will be added to the Hamiltonian $H_\lambda$, making the problem in general map to a nonlinear and nonlocal time-independent Schrödinger equation.

Since Eq.~\eqref{eq:schrodinger} is a linear, time-independent Schrödinger equation, many analytic solutions exist and are well documented. Indeed, examples studied in this \type map onto quantum mechanics problems such as the two-state system (bit flip control), the quantum harmonic oscillator (Mexican--hat potential, by minimizing mean-squared distance instead of maximizing its inverse), and the quantum rotor~\cite{hradil2006minimum} (microbial navigation), for which analytical solutions for the ground state and energy are well-known. Even when analytic solutions are not possible, the linear time-independent Schrödinger equation has been thoroughly studied in the literature, and numerical techniques for efficiently computing ground states and energies are abundant. Solutions to the nonlinear and nonlocal version are generally far less tractable, but numerical techniques nonetheless exist.

The ground-state energy $E_0(\lambda)$, once determined, fully characterizes the performance-information Pareto frontier. Taking the derivative of $H_\lambda$ with respect to $\lambda$, multiplying by $\psi^*(\x)$, and integrating over $\x$, we find that the performance is given by the derivative $E_0'$ with respect to the Lagrange multiplier $\lambda$,
\begin{equation}
\mathcal{P}[\rho] = -E_0'(\lambda).
\end{equation}
Moreover, the minimum information rate $\dot{\mathcal{T}}_{\x\to\dxc}$, equal to the passive entropy rate $\dot{S}_p$ can then be written in terms of the ground-state energy as 
\begin{equation}
\dot{\mathcal{T}}_{\x\to\dxc} = E_0(\lambda) - \lambda E_0'(\lambda)\equiv B_0(\lambda),
\end{equation}
where we have defined the function $B_0(\lambda)$ in terms of $E_0$ and $E_0'$. The Pareto frontier between performance and information is then, in full generality,
\begin{equation}\label{eq:quantumfrontier}
\mathcal{P}[\rho]= -E_0'\left(B_0^{-1}\left[\dot{\mathcal{T}}_{\x\to\dxc}\right]\right).
\end{equation}

%%%%%%%%%%%%%%%%%%%%%%%%%%%%%%%%%%%%%%%%%%%%%%%%%%%%%%%%%%%%%%%
\section{Time-Reversal Control Protocol}\label{app:optimal_protocol}
Here we construct a control protocol which saturates our bound~\eqref{eq:main_inequality} in the case where the passive dynamics are state-independent, i.e., $\de\x_t^p$ is independent of $\x_t$. For the passive dynamics in Eq.~\eqref{General_SDE}, this assumption would require a passive update of the form
\begin{equation}\label{eq:stateindependentpassivedynamics}
\de\x^p_t = \bm{F}\de t + \sqrt{2\diffusivity}\,\de\bm{W}^p_t + \Delta\x\, \mathrm{d}J(\lambda(\Delta\x)),
\end{equation}
for constant $\bm{F}$ and rate $\lambda(\Delta\x)$ independent of $\x$. We assume that the state $\x_t$ is fully observable, and that there are no restrictions on the actions available to the controller. 

We begin by noting that two inequalities were used in deriving Eq.~\eqref{eq:main_inequality}. The first was the data-processing inequality,
\begin{equation}
I[\y_t;\{\dxc\}_{-\infty}^t]\geq I[\x_{t+\de t};\{\dxc\}_{-\infty}^t].
\end{equation}
The second was 
\begin{equation}
I[\dxc_t;\x_t|\{\dxc\}_{-\infty}^{t-\de t}]\leq I[\dxc_t;\{\x\}_{-\infty}^t|\{\dxc\}_{-\infty}^{t-\de t}].
\end{equation}
Accounting for the difference between the left- and right-hand sides of these two inequalities, we find that the difference between the transfer entropy and passive entropy increase satisfies the equality
\begin{equation}\label{eq:inequality_gap}
\Delta\mathcal{T}_{\x\to\dxc} - \Delta S_p = \underbrace{I[\dxc_t;\{\x\}_{-\infty}^{t-\de t}|\x_t,\{\dxc\}_{-\infty}^{t-\de t}]}_{\mathrm{Term\,1}} + \underbrace{I[\y_t;\{\dxc\}_{-\infty}^{t}|\x_{t+\de t}]}_{\mathrm{Term\,2}}.
\end{equation}
Term 1 is equal to zero if and only if the control update does not depend on the history of $\x$ given the current state $\x_t$ and past control history. This condition is satisfied if
\begin{equation}
p(\dxc_t|\x_t,\{\x\}_{-\infty}^{t-\de t},\{\dxc\}_{-\infty}^{t-\de t}) = p(\dxc_t|\x_t,\{\dxc\}_{-\infty}^{t-\de t}).
\end{equation}
Term 2 is zero if and only if the previous control target $\y_t$ is independent of the control history $\{\dxc\}_{-\infty}^t$ given $\x_{t+\de t}$. This is achieved if
\begin{equation}\label{eq:term2condition}
p(\y_t|\x_{t+\de t},\{\dxc\}_{-\infty}^t) = p(\y_t|\x_{t+\de t}).
\end{equation}

We now construct a control protocol that satisfies these two conditions. First note that, at steady state, $\x_t$ and $\x_{t+\de t}$ both follow the distribution $\rho(\x)$. The variable $\y_t$, meanwhile, follows a (generally different) distribution $\rho'(\y)$. Defining the passive transition kernel
\begin{equation}
K_p(\x'|\y) \equiv p(\x_{t+\de t}|\y_t),
\end{equation}
these two distributions are related as
\begin{equation}
\rho(\x') = \int\de\y K_p(\x'|\y)\rho'(\y).
\end{equation}
Now define the reverse transition kernel,
\begin{equation}
R_p(\y|\x') \equiv p(\y_t|\x_{t+\de t}).
\end{equation}
By Bayes' rule, this must satisfy
\begin{equation}
R_p(\y|\x') = \frac{\rho'(\y)K_p(\x'|\y)}{\rho(\x')}.
\end{equation}
Finally, define the active transition kernel as
\begin{equation}
K_c(\y|\x) \equiv p(\y_t|\x_t),
\end{equation}
and set it equal to the passive reverse kernel 
\begin{equation}
K_c(\y|\x) = R_p(\y|\x).
\end{equation}
This protocol depends only on the current state $\x_t$, so sets Term 1 to zero. 

To see that Term 2 is also set to zero by this protocol, note that for state-independent passive dynamics the passive kernel takes the form $K_p(\x'|\y)=q(\x'-\y)$ for some function $q$. To do so, we will first show that the distribution $p(\x_{t+\de t}|\{\de\x^c\}_{-\infty}^t) = \rho(\x_{t+\de t})$ does not depend on the control history, doing so by induction. Let us suppose first that $p(\x_{t}|\{\de\x^c\}_{-\infty}^{t-\de t}) = \rho(\x_{t})$. We then compute
\begin{equation}
\begin{aligned}
p(\y_t,\x_{t+\de t},\de\x^c_t|\{\de\x^c\}_{-\infty}^{t-\de t}) & = p(\x_t=\y_t-\de\x^c_t|\{\de\x^c\}_{-\infty}^{t-\de t}) \cdot K_c(\y_t|\x_t)\cdot K_p(\x_{t+\de t}|\y_t)\\
& = \rho(\y_t-\de\x^c_t) \cdot \frac{\rho'(\y_t)q(\x_{t}-\y_t)}{\rho(\x_t)}\cdot q(\x_{t+\de t}-\y_t)\\
& = \rho'(\y_t) q(-\de\x^c_t) q(\x_{t+\de t}-\y_t).
\end{aligned}
\end{equation}
Integrating over $\y_t$ and $\x_{t+\de t}$ then gives $p(\de\x^c_t|\{\de\x^c\}_{-\infty}^{t-\de t}) = q(-\de\x^c_t)$. We can then write
\begin{equation}
p(\y_t,\x_{t+\de t}|\{\de\x^c\}_{-\infty}^{t}) = \frac{p(\y_t,\x_{t+\de t},\de\x^c_t|\{\de\x^c\}_{-\infty}^{t-\de t})}{p(\de\x^c_t|\{\de\x^c\}_{-\infty}^{t-\de t})} = \rho'(\y_t)q(\x_{t+\de t}-\y_t).
\end{equation}
Integrating over $\y_t$ gives $p(\x_{t+\de t}|\{\de\x^c\}_{-\infty}^{t}) = \rho(\x_{t+\de t})$, thus proving that $\x_t$ being independent of $\{\de\x^c\}_{-\infty}^{t}$ is self-consistent and true by induction. Finally, computing
\begin{equation}
p(\y_t|\x_{t+\de t},\{\de\x^c\}_{-\infty}^{t}) = \frac{p(\y_t,\x_{t+\de t}|\{\de\x^c\}_{-\infty}^{t})}{p(\x_{t+\de t}|\{\de\x^c\}_{-\infty}^{t})} = \frac{\rho'(\y_t)q(\x_{t+\de t}-\y_t)}{\rho(\x_{t+\de t})} = R_p(\y_t|\x_{t+\de t})
\end{equation}
shows that Eq.~\eqref{eq:term2condition} is satisfied, and thus Term 2 is zero.

Thus an optimal control protocol for state-independent passive dynamics is the Bayesian time reversal of the passive dynamics. The protocol samples the post-control state $\y$ from the posterior distribution of where the passive dynamics must have come from to arrive at the pre-control state $\x_t$. We can evaluate the control dynamics explicitly given the Markovian passive dynamics posited in Eq.~\eqref{eq:stateindependentpassivedynamics}. A simple calculation gives:
\begin{equation}
\dxc_t = \left[-\bm{F} + 2\diffusivity \nabla\ln \rho(\x_t)\right]\de t + \sqrt{2\diffusivity}\de\bm{W}_t^c + \Delta\x \de J\left(\frac{\lambda(-\Delta\x)\rho(\x_t+\Delta\x)}{\rho(\x_t)}\right).
\end{equation}
Thus given a steady-state distribution $\rho(\x)$, probabilistic time reversal gives an explicit optimal control protocol for state-independent passive dynamics.

We conjecture, and show in the main text through examples, that probabilistically time-reversing the passive dynamics yields a near-optimal protocol more generally, approaching the bound~\eqref{eq:main_inequality} in the high-information limit. Here we derive an analytic upper bound on the information rate required to implement the time-reversal protocol, allowing quantitative estimation of the protocol's distance from the bound. For arbitrary passive dynamics as given by Eq.~\eqref{General_SDE}, the time-reversal protocol is
\begin{equation}\label{eq:generaltimereversalprotocol}
\dxc_t = \left[-\bm{F}(\x_t) + 2\diffusivity \nabla\ln \rho(\x_t)\right]\de t + \sqrt{2\diffusivity}\de\bm{W}_t^c + \Delta\x \de J\left(\frac{\lambda(\x_t + \Delta\x,-\Delta\x)\rho(\x_t+\Delta\x)}{\rho(\x_t)}\right).
\end{equation}
This protocol depends only on the current state $\x_t$, so the instantaneous control information rate $\dot{I}_c^+$ is an upper bound on the full-history transfer entropy rate $\dot{\mathcal{T}}_{\x\to\de\x^c}$ (Eq.~\eqref{eq:controlinfoupperbound}). Thus, the difference between the information rate $\dot{\mathcal{T}}_{\x\to\de\x^c}$ and the bound $\dot{S}_p$ is upper-bounded by
\begin{equation}\label{eq:timereversalUB}
\begin{aligned}
\dot{\mathcal{T}}_{\x\to\de\x^c}-\dot{S}_p & \leq \dot{I}^+_c-\dot{S}_p\\
&= \frac{1}{4}\left[\left\langle\bm{F}\cdot \diffusivity^{-1}\bm{F}\right\rangle_\rho - \left\langle\bm{F}\right\rangle_\rho\cdot\diffusivity^{-1}\left\langle \bm{F}\right\rangle_\rho\right] + \iint\de\x\de\Delta\x\, \rho(\x)\lambda(\x,\Delta\x)\ln\frac{\lambda(\x,\Delta\x)}{\bar{\lambda}(\Delta\x)},
\end{aligned}
\end{equation}
where we used Eqs.~\eqref{eq:explicitpassiveentropy} and \eqref{eq:localcontrolinfoevaluated}, and defined $\bar{\lambda}(\Delta\x)\equiv \langle\lambda(\x,\Delta\x)\rangle_\rho$.

%%%%%%%%%%%%%%%%%%%%%%%%%%%%%%%%%%%%%%%%%%%%%%%%%%%%%%%%%%%%%%
\section{Controlling a Brownian Particle in a Harmonic Trap}\label{app:harmonictrap}
As an additional example beyond those covered in the main text, we consider the problem of controlling a Brownian particle in a harmonic trap, with passive dynamics (in dimensionless units)
\begin{equation}
\de x^p_t = -x_t\de t + \sqrt{2}\de W^p_t.
\end{equation}
This analysis can be mapped onto the problem studied previously in Ref.~\cite{lestas2010fundamental}, our results in this section are consistent with theirs.
Here the goal of the controller is to achieve a target variance $s=\mathrm{Var}(x)$ using the minimum required information rate $\dot{\mathcal{T}}_{x\to\de x_c}$. We stipulate that the controller apply a noisy linear force, as
\begin{equation}
\de x^c_t = -k_c x_t\de t + \sqrt{2D_c}\de W^c_t.
\end{equation}
For fixed $s$, the control protocol has only one degree of freedom, which we take to be $k_c$. The controller noise is then $D_c = s(1+k_c) - 1$.

\begin{figure*}[t]
    \centering
    \includegraphics[width=0.8\linewidth]{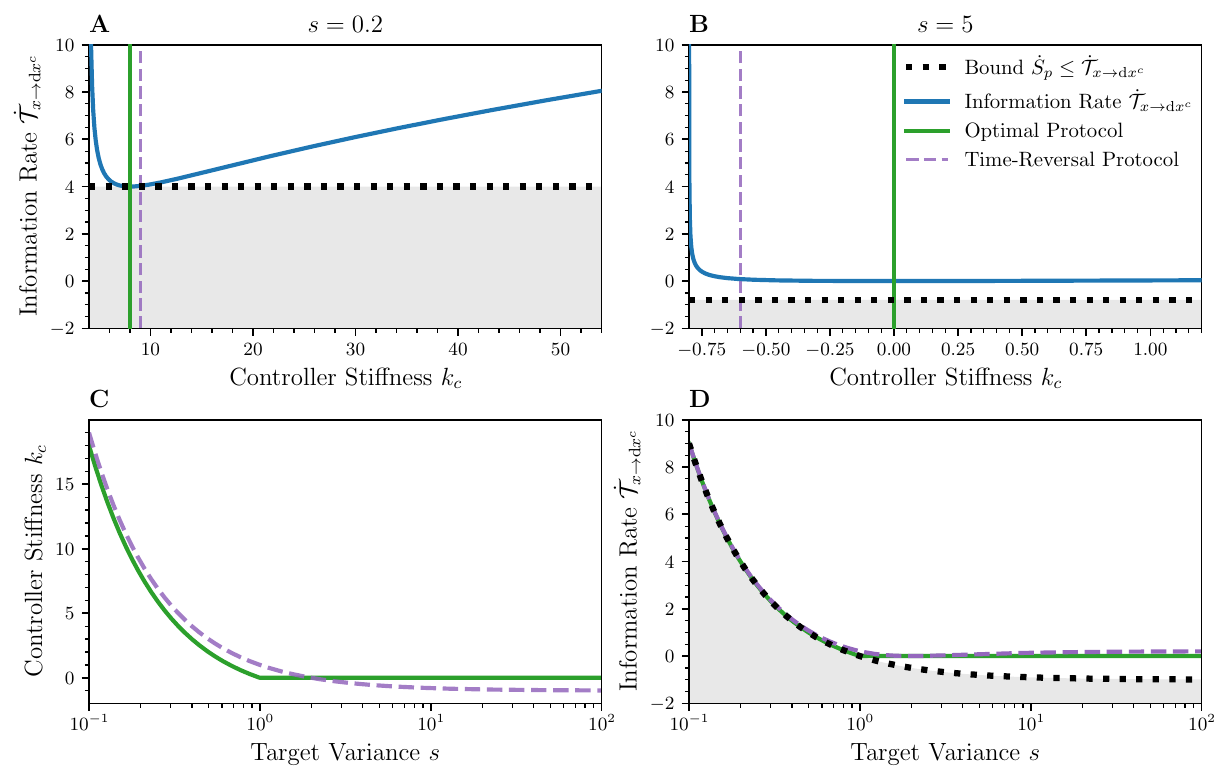}
    \caption{Control of a Brownian particle in a harmonic trap. (a) and (b) Information rate (blue) as a function of controller stiffness $k_c$ for target variance $s=0.2$ (a) and $s=5$ (b). Black dotted line: bound, Eq.~\eqref{eq:main_inequality}. Vertical lines: optimal protocol (green) and time-reversal protocol (purple). (c) Controller stiffness as a function of target variance $s$ for the optimal (green) and time-reversal (purple) protocols. (d) Information rate vs target variance. Black: bound, Eq.~\eqref{eq:main_inequality}. Green: optimal protocol. Purple: time-reversal protocol. }
    \label{fig:harmonictrap}
\end{figure*}

The passive entropy rate is $\dot{S}_p = 1/s-1$, which in turn gives a lower bound on the information rate from Eq.~\eqref{eq:main_inequality}. Intriguingly, we see that when $s>1$, i.e., the target variance is larger than the equilibrium variance of the passive dynamics alone, the passive entropy rate becomes negative. Since the information rate must be nonnegative, for $s>1$ the bound is thus no longer tight. We can nonetheless compute the information rate,
\begin{equation}
\dot{\mathcal{T}}_{x\to\de x_c} = \frac{1}{2}\left[\sqrt{1 + \frac{k_c^2}{s(1+k_c) - 1}}-1\right],
\end{equation}
and minimize it with respect to the control stiffness $k_c$ to obtain an optimal protocol given by $k_c = 2\max\{0,1/s - 1\}$. The time-reversal protocol, meanwhile, is $k_c = 2/s-1$. While these protocols are not equal, the time-reversal protocol is nonetheless near-optimal, and requires the same information rate in the high-information (or equivalently low-$s$) limit [Fig.~\ref{fig:harmonictrap}].

%%%%%%%%%%%%%%%%%%%%%%%%%%%%%%%%%%%%%%%%%%%%%%%%%%%%%%%%%%%%%%
\section{Details for Bit Flip Control}\label{app:bitflipcontrol}
Here we provide expanded details and derivations for the bit flip control example in the main text. This application consists of controlling a single bit (with $\x\in\{0,1\}$), which passively flips its state at rate $\gamma$, to maximize the probability $\rho_0$ of it being in state $0$. The passive entropy rate is
\begin{equation}
\dot{S}_p = (2\rho_0-1)\gamma\ln\frac{\rho_0}{1-\rho_0}.
\end{equation}
This straightforwardly gives the Pareto frontier between the performance metric $\mathcal{P}[\rho] = \rho_0$ and the information rate,
\begin{equation}
\dot{\mathcal{T}}_{x\to\de x^c} \geq (2\rho_0-1)\gamma\ln\frac{\rho_0}{1-\rho_0}.
\end{equation}

Given $\gamma$, the controller can choose the control jump rates $\lambda_0$ and $\lambda_1$ to achieve a target performance $\mathcal{P}[\rho] = \rho_0$ with limited information. For mathematical convenience, we take the controller action to be addition by one modulo two. To enforce a steady-state, the rates and probabilities must satisfy
\begin{equation}
\rho_0(\gamma+\lambda_0)= (1-\rho_0)(\gamma+\lambda_1).
\end{equation}
We can thus write $\lambda_1$ in terms of $\gamma,\rho_0$, and $\lambda_0$ as
\begin{equation}\label{eq:bitflip_lambda1}
\lambda_1 = \frac{\rho_0}{1-\rho_0}(\gamma+\lambda_0) - \gamma.
\end{equation}

The optimal control protocol~\eqref{eq:timereversalcontrol} which probabilistically time-reverses the passive dynamics is
\begin{equation}
\lambda_0 = \gamma\frac{1-\rho_0}{\rho_0}.
\end{equation}

To explore the information rates of suboptimal protocols, we compute the information rate $\dot{\mathcal{T}}_{\x\to\dxc}$. To do this, we define the probability $q_t \equiv p(x_t=1|\{\dxc\})$, which we can think of as a time-evolving belief about the current state of the bit. We also define the expected rate of controlled jumps given the controller history, $\ell_t(q) \equiv \mathbb{E}[\lambda_x|\{\dxc\}] = (1-q_t)\lambda_0 + q_t\lambda_1$. Using Eq.~\eqref{eq:SITEdef}, we have
\begin{equation}\label{eq:bitflipTEcalc}
\dot{\mathcal{T}}_{\x\to\dxc} = \left\langle \lambda_x \ln\frac{\lambda_x}{\ell(q)}\right\rangle_{p(q)}.
\end{equation}
Evaluating this average requires computing the steady-state average $p(q)$, which we will do by constructing a differential equation describing the dynamics of $q_t$. First consider the time between discrete controller actions. When no controller action takes place, the passive dynamics push the belief $q_t$ towards $1/2$ at rate
\begin{equation}
\dot{q} = \gamma(1-2q).
\end{equation}
The absence of a controller action is also informative. Since $\lambda_1\geq\lambda_0$ (as required for $\rho_0\geq1/2$), this absence of a control action updates the belief towards $x=0$ at rate 
\begin{equation}
\dot{q}=-(\lambda_1-\lambda_0)q(1-q).
\end{equation}
Finally, controlled jumps in state provide discrete updates to the belief. At the instantaneous moment of a controlled jump, the post-jump belief $p(x^+_t|\mathrm{\,jump\, at \,}t)$ must satisfy 
\begin{equation}
q^+_t = p(x^+_t|\mathrm{controlled\,jump\, at \,}t) = \frac{p(\mathrm{controlled\,jump\, at \,}t\,|x^+_t) p(x^+_t)}{p(\mathrm{controlled\,jump\, at \,}t)} = \frac{\lambda_0(1-q^-_t)}{\ell_t(q^-_t)}\equiv R(q^-_t).
\end{equation}
Here the notation $q^+_t = R(q^-_t)$ means that $q^+_t$ is the post-jump belief given a controlled jump at time $t$ and pre-jump belief $q^-_t$. Combining all of this, we obtain the following Fokker--Planck dynamics for $p(q)$:
\begin{equation}
\partial_t p_t(q) = -\partial_q\left[\underbrace{\left(\gamma(1-2q) - (\lambda_1 - \lambda_0)q(1-q)\right)}_{f(q)}p_t(q)\right] - \ell(q)p_t(q) + \int \de y\, \ell(y)p_t(y)\delta(q - R(y)).
\end{equation}
We can simplify this significantly by noting that the function $R$ is its own inverse, $R(R(q))=q$, and using the delta function identity $\delta(q-R(y)) = \delta(y-R(q))/|R'(R(q))|$, to evaluate the integral
\begin{equation}
\int\de y\, \ell(y)p_t(y)\delta(q - R(y)) = \frac{(\lambda_0\lambda_1)^2}{\ell(q)^3}p_t(R(q)).
\end{equation}
Thus the final equation defining the steady-state belief probability $p(q)$ is
\begin{equation}
\partial_q\left[f(q)p(q)\right] + \ell(q)p(q) = \frac{(\lambda_0\lambda_1)^2}{\ell(q)^3}p(R(q)).
\end{equation}
Solving numerically for $p(q)$ is then sufficient to calculate the information rate using Eq.~\eqref{eq:bitflipTEcalc}.

%%%%%%%%%%%%%%%%%%%%%%%%%%%%%%%%%%%%%%%%%%%%%%%%%%%%%%%%%%%%%%
\section{Details for Mexican--hat Potential}\label{app:mexicanhat}
Here we provide details for calculating the Pareto frontiers and optimal control protocols for the 2D Brownian particle in a nonlinear potential $V(x)$. The passive dynamics (in scaled units) are
\begin{equation}
\de\x_t^p = -\nabla V(x) \de t + \sqrt{2}\,\de\bm{W}^p_t.
\end{equation}
For a given steady state $\rho(\x)$, the passive entropy rate is
\begin{equation}
\dot{S}_p = \mathcal{I}[\rho] - \left\langle\nabla^2V(\x)\right\rangle_\rho,
\end{equation}
where $\mathcal{I}[\rho]$ is the Fisher information with respect to position shifts, \begin{equation}
\mathcal{I}[\rho] \equiv \left\langle\left(\nabla\ln \rho(\x)\right)\cdot\nabla\ln \rho(\x)\right\rangle_\rho.
\end{equation} 

The goal of the feedback controller is to tightly localize the system near some desired point $\x^*$, maximizing performance $\mathcal{P}[\rho] = 1/\left\langle |\x-\x^*|^2\right\rangle_\rho$. When the particle is tightly localized, approximating the potential energy landscape as locally quadratic using the Hessian $H_* = \nabla^2 V(\x^*)$ gives
\begin{equation}
\dot{\mathcal{T}}_{\x\to\de\x^c}\geq\dot{S}_p = \sum_{\lambda_i}\mathrm{Re}(\lambda_i) +\mathcal{I}[\rho],
\end{equation}
where $\{\lambda_i\}$ are the eigenvalues of $\bm{A} = -H_*$.

The form above is reminiscent of the Data Rate Theorem (DRT)~\cite{tatikonda2004control}, which states that to control an unstable linear system $\dot{\x} = \bm{A}\x$, the minimum information required is the sum of the real parts of the eigenvalues of $\bm{A}$ with positive real part~\cite{tatikonda2004control,nair2004topological,nair2007feedback}:
\begin{equation}\label{eq:drt}
\dot{I}\geq \sum_{\mathrm{Re}(\lambda_i)>0}\mathrm{Re}(\lambda_i).
\end{equation}
In the linear limit, we can connect our bound~\eqref{eq:main_inequality} to the DRT by first projecting the state $\x$ onto only the unstable directions to obtain a projected state $\z$, then restricting the controller to only act on $\z$ with control update that likewise depends only on $\{\z\}_{-\infty}^t$. Doing so yields a generalization of the DRT which quantifies the added information cost to counteract passive diffusion when controlling a passively unstable linear system:
\begin{equation}\label{eq:drt_stochastic}
\dot{\mathcal{T}}_{\z\to\de\z^c} \geq \sum_{\mathrm{Re}(\lambda_i)>0}\mathrm{Re}(\lambda_i) + \mathcal{I}[\rho(\z)].
\end{equation}

As an illustrative example, consider the two-dimensional Mexican--hat potential (in scaled units),
\begin{equation}
V(r) = -\frac{1}{2}r^2 + \frac{1}{4}r^4,\; r=\sqrt{x^2+y^2}.
\end{equation}
The origin $r=0$ is an unstable fixed point, while the ring with $r=1$ is a set of stable fixed points. We consider the control problem of minimizing the mean-squared distance from the fixed points $\x^*_1 = (0,0)$, and $\x^*_2 = (1,0)$.

From Eq.~\eqref{eq:explicitpassiveentropy}, the passive entropy rate is
\begin{equation}
\dot{S}_p = \mathcal{I}[\rho] + 2 - 4\langle r^2\rangle_\rho;
\end{equation}
this in turn sets the general lower bound on the information rate required for control. We will determine Pareto frontiers using the Schrödinger equation mapping from App.~\ref{app:schrodinger_mapping}. The Schrödinger equation corresponding to the control performance extremization, applied to this system, is
\begin{equation}
\left[-4\nabla^2 + 2 - 4 r^2 -\lambda |\x-\x^*|^2\right]\psi(\x) = E_0(\lambda)\psi(x).
\end{equation}
For mathematical convenience, we consider minimization of $\langle|\x-\x^*|^2\rangle_\rho$ rather than maximization of its inverse.

We begin with the first target $\x^*_1$. The Schrödinger equation is then
\begin{equation}
\left[-4\nabla^2 + 2 - (4+\lambda) r^2\right]\psi(\x) = E_0(\lambda)\psi(x).
\end{equation}
This is a standard isotropic 2D quantum harmonic oscillator, and thus has ground-state energy 
\begin{equation}
E_0(\lambda) = 2 + 4\sqrt{-\lambda-4}.
\end{equation}
This requires $\lambda<-4$.
Using the formula for the Pareto frontier~\eqref{eq:quantumfrontier}, we obtain
\begin{equation}
\left\langle |\x-\x^*|^2\right\rangle_\rho^{-1}\leq \frac{1}{8} \left(\dot{\mathcal{T}}_{\x\to\dxc} - 2 + \sqrt{\dot{\mathcal{T}}_{\x\to\dxc}^2 - 4\dot{\mathcal{T}}_{\x\to\dxc} + 68}\right).
\end{equation}
The ground-state probability density, meanwhile, is
\begin{equation}
\rho(\x) =\frac{\sqrt{-\lambda-4}}{2\pi}\exp\left(-\frac{\sqrt{-\lambda-4}}{2}\left[\x - \x_1^*\right]^2\right).
\end{equation}

We then consider the second target $\x^*_2$. The Schrödinger equation is
\begin{equation}
\left[-4\nabla^2 + 2 - 4r^2 - \lambda|\x-\x^*_2|^2\right]\psi(\x) = E_0(\lambda)\psi(x).
\end{equation}
Using $\x=(x,y)$ and completing the square, we can rewrite this as
\begin{equation}
\begin{aligned}
E_0(\lambda)\psi(x) & = \left[-4\nabla^2 + 2 - 4(x^2+y^2) - \lambda\left(((x-1)^2 + y^2\right)\right]\psi(\x)\\
& = \left[-4\nabla^2 + 2 - (4+\lambda)\left[\left(x-\frac{\lambda}{4+\lambda}\right)^2 + y^2\right] - \lambda + \frac{\lambda^2}{4+\lambda}\right]\psi(\x).
\end{aligned}
\end{equation}
The ground-state energy is then
\begin{equation}
E_0(\lambda) = 2 + \frac{\lambda^2}{4+\lambda} - \lambda + 4\sqrt{-\lambda - 4},
\end{equation}
while the probability density is
\begin{equation}
\rho(\x) = \frac{\sqrt{-\lambda-4}}{2\pi}\exp\left(-\frac{\sqrt{-\lambda-4}}{2}\left[\left(x-\frac{\lambda}{4+\lambda}\right)^2 + y^2\right]\right).
\end{equation}
Here we see the distribution is offset slightly from the target $x=1$, approaching it only asymptotically as $\lambda\to-\infty$.

%%%%%%%%%%%%%%%%%%%%%%%%%%%%%%%%%%%%%%%%%%%%%%%%%%%%%%%%%%%%%%
\section{Details for Microbial Navigation}\label{app:navigation}
Here we provide details for the microbial navigation application considered in the main text. We model chemotaxis as constant-speed motion (with speed $v_0$) at a heading angle $\theta\in[-\pi,\pi)$ with respect to the gradient direction. The angle $\theta$ is both the signal acquired by the sensor and the variable on which the controller acts, obeying overdamped Langevin dynamics on the circle:
\begin{equation}
\de\theta_t =\! \sqrt{2D_r}\,\de W^p_t + \de \theta^c_t.
\end{equation}
The passive dynamics are rotational diffusion with coefficient $D_r$, such that the passive entropy rate is $\dot{S}_p=D_r\mathcal{I}[\rho(\theta)]$. The performance metric, meanwhile, is the up-gradient velocity, $\mathcal{P}[\rho] = v/v_0=\langle\cos\theta\rangle_\rho$.

\subsection{Optimal Control Protocol with Full Observability}
For the navigation problem, the corresponding Schrödinger equation is
\begin{equation}
\left[-4D_r\partial_\theta^2 -\lambda\cos\theta\right]\psi(\theta) = E_0(\lambda)\psi(\theta).
\end{equation}
This is the Hamiltonian for a quantum rotor which has been studied previously~\cite{hradil2006minimum}. The ground-state distribution is
\begin{equation}
\rho(\theta) = \frac{1}{2\pi}\mathrm{ce}_0(\theta/2,-\lambda/2D_r)^2,
\end{equation}
with corresponding ground-state energy $D_r a_0(-\lambda/2D_r)$, where $\mathrm{ce}_0$ is the zeroth order Mathieu C function of the first kind~\cite{arfken2005mathematical}, and $a_0$ is the zeroth order Mathieu characteristic function. 

From Eq.~\eqref{eq:timereversalcontrol}, the optimal control force is
\begin{equation}
F_c(\theta)=4D_r \frac{\partial_\theta \mathrm{ce}_0\left(\theta/2,b_0^
{-1}\left(\dot{\mathcal{T}}_{\theta\to\de\theta^c}/D_r\right)\right)}{\mathrm{ce}_0\left(\theta/2,b_0^
{-1}\left(\dot{\mathcal{T}}_{\theta\to\de\theta^c}/D_r\right)\right)}.
\end{equation} 

Employing this optimal protocol leads to a velocity--information Pareto frontier:
\begin{equation}\label{eq:navigationfrontier}
v/v_0 \leq \frac{1}{2}a_0'\left[b_0^
{-1}\left(\dot{\mathcal{T}}_{\theta\to\de\theta^c}/D_r\right)\right],
\end{equation}
where $a_0(x)$ is the Mathieu characteristic function (and $a_0'(x)$ its derivative), and $b_0(x) = a_0(x) - x a_0'(x)$ [Fig.~\ref{fig:navigation}]. This shows that the velocity--information Pareto frontier obtained in our previous work~\cite{betancourt2026discrete} applies to the full history-dependent transfer entropy rate, thus generalizing to arbitrary control protocols with memory. At low information this frontier scales as $v/v_0\sim\sqrt{ \dot{\mathcal{T}}_{\theta\to\de\theta^c}/2D_r}$, and at high information scales as $v/v_0\sim 1- D_r/2\dot{\mathcal{T}}_{\theta\to\de\theta^c}$. 

\subsection{Run-and-Tumble Navigation}
Now consider a Run-and-Tumble strategy, where the only action available is to fully reorient the heading with rate $\lambda(\{\theta\}_{-\infty}^t,\{\de\theta^c\}_{-\infty}^t)$ such that the angle change is uniformly distributed on $[-\pi,\pi)$. We consider the Markovian version of this strategy where $\lambda = \lambda(\theta)$. Here we consider the control action to be initiation of a tumble, rather than the specific angular change induced.

The transfer entropy rate can be computed exactly for this strategy. To do so, we note that the entire history of the state and controller are fully characterized by the current heading $\theta$ and the time $\tau$ since the last tumble. Because the passive dynamics are Markovian and the tumbles fully reset the heading, no other history carries further information. Thus we need only determine the distribution $q(\theta,\tau)$, which is the probability that a run still survives at time $\tau$ and the current heading is $\theta$. Integrating over $\theta$ and $\tau$ does not equal one, so this is not a normalized joint probability distribution. Integrating over $\theta$ gives the survival function
\begin{equation}
S(\tau) = \int \de\theta q(\theta,\tau),
\end{equation}
which is the probability of a run having survived without interruption at time $\tau$. Integrating this over $\tau$ in turn gives the mean run time,
\begin{equation}
\bar{\tau} = \int_0^\infty S(\tau)\de\tau.
\end{equation}
The normalized probability distribution over run time $\tau$ and heading $\theta$ is $p(\theta,\tau) = q(\theta,\tau)/\bar{\tau}$. The distribution $q(\theta,\tau)$ evolves according to
\begin{equation}
\partial_\tau q(\theta,\tau) = D_r\partial_\theta^2 q(\theta,\tau) - \lambda(\theta)q(\theta,\tau),
\end{equation}
with initial condition $q(\theta,0)=1/2\pi$. This makes it clear it is not a normalized joint probability. This PDE is linear and thus easily solvable numerically given $\lambda(\theta)$.

The conditional distribution over the heading given a time $\tau$ since the last tumble is $\rho(\theta|\tau) = q(\theta,\tau)/S(\tau)$. We then define $\bar\lambda(\tau) = \int\de\theta \rho(\theta|\tau)\lambda(\theta)$ as the mean tumble rate given time $\tau$ since the last tumble. It should also satisfy
\begin{equation}
\de_\tau S(\tau) = -\bar{\lambda}(\tau)S(\tau),
\end{equation}
with $S(0)=1$.

The transfer entropy rate, given that the control strategy is Markovian, is computed as
\begin{equation}
\begin{aligned}
\dot{\mathcal{T}}_{\theta\to\de\theta_c} & = \left\langle\lambda(\theta)\ln\frac{\lambda(\theta)}{\mathrm{E}[\lambda(\theta)|\{\de\theta^c\}_{-\infty}^t]}\right\rangle_{p(\theta,\{\de\theta^c\}_{-\infty}^t)}\\
& = \left\langle\lambda(\theta)\ln\frac{\lambda(\theta)}{\bar{\lambda}(\tau)}\right\rangle_{p(\theta,\tau)}.
\end{aligned}
\end{equation}

\subsection{Optimal Protocol with Limited Observability}
Run-and-Tumble is not the optimal control protocol using only measurements of $\cos\theta$ in the low-information limit. Here we prove that Run-Reverse is the optimal control protocol in the low-information limit, even allowing the use of arbitrary memory, given that only $\cos\theta$ can be measured. For mathematical simplicity, we will consider control using only measurements of $|\theta|$ which is equivalent to using only $\cos\theta$. 

First, note that as shown above, maximizing up-gradient velocity $v/v_0=\langle\cos\theta\rangle$ given fixed passive entropy rate $\dot{S}_p$ requires a steady-state distribution
\begin{equation}
\rho(\theta) = \frac{1}{2\pi}\mathrm{ce}_0\left[\theta/2,b_0^
{-1}\left(\dot{\mathcal{T}}_{\theta\to\de\theta^c}/D_r\right)\right]^2.
\end{equation}
To saturate the bound, we must then find a control protocol that achieves this distribution using only measurements of $|\theta|$, while also saturating the bound~\eqref{eq:main_inequality}.

Expanding $\rho(\theta)$ in the limit of small information rate, $\epsilon\equiv \left(\dot{\mathcal{T}}_{\theta\to\de\theta^c}/D_r\right)^{1/2}\ll1$, gives
\begin{equation}
\rho(\theta) = \frac{1}{2\pi}\left[1 +\sqrt{2}\,\epsilon\cos\theta \right] + \mathcal{O}\left(\epsilon^2\right).
\end{equation}
In this same limit, the passive entropy rate is
\begin{equation}
\dot{S}_p = \epsilon^2 D_r + \mathcal{O}\left(\epsilon^3\right).
\end{equation}
We similarly expand the control protocol around zero information as
\begin{equation}
\lambda\left(\Delta\theta|\{|\theta|\}_{-\infty}^{t},\{\de\theta^c\}_{-\infty}^{t-\de t}\right) = r\left(\Delta\theta|\{\de\theta^c\}_{-\infty}^{t-\de t}\right) + \epsilon\,\eta\left(\Delta\theta|\{|\theta|\}_{-\infty}^{t},\{\de\theta^c\}_{-\infty}^{t-\de t}\right) + \mathcal{O}\left(\epsilon^2\right).
\end{equation}
Here $\eta$ and $r$ are functions which do and do not respectively depend on the state $\theta$ and its history.
Note that using arbitrary $\lambda(\Delta\theta,|\theta|)$ can describe any arbitrary additive jump or drift-diffusion control protocol~\cite{betancourt2026discrete}. When $\epsilon=0$ this protocol has zero information rate.

Now consider the gap in the bound~\eqref{eq:main_inequality}, which we showed in App.~\ref{app:optimal_protocol} can be written as
\begin{equation}
\Delta\mathcal{T}_{\theta\to\de \theta^c} - \Delta S_p = \underbrace{I[\de\theta^c_t;\{\theta\}_{-\infty}^{t-\de t}|\theta_t,\{\de\theta^c\}_{-\infty}^{t-\de t}]}_{\mathrm{Term\,1}} + \underbrace{I[y_t;\{\de\theta^c\}_{-\infty}^{t}|\theta_{t+\de t}]}_{\mathrm{Term\,2}}.
\end{equation}
Since $\dot{S}_p\sim\epsilon^2$ to leading order, saturating the bound in the low-information limit requires that both term 1 and term 2 scale as $\epsilon^3$. 

We consider term 1 explicitly. Defining $\bar\eta(\Delta\theta,|\theta|_t,\{\de\theta^c\}_{-\infty}^{t-\de t}) = \left\langle\left.\eta\left(\Delta\theta,\{|\theta|\}_{-\infty}^{t},\{\de\theta^c\}_{-\infty}^{t-\de t}\right)\right||\theta|_t,\{\de\theta^c\}_{-\infty}^{t-\de t}\right\rangle$, the leading-order term is
\begin{equation}
\frac{1}{\de t}I[\de\theta^c_t;\{\theta\}_{-\infty}^{t-\de t}|\theta_t,\{\de\theta^c\}_{-\infty}^{t-\de t}] = \frac{\epsilon^2}{2}\left\langle \int\de\Delta\theta \frac{\left[\eta\left(\Delta\theta,\{|\theta|\}_{-\infty}^{t},\{\de\theta^c\}_{-\infty}^{t-\de t}\right)-\bar\eta(\Delta\theta,|\theta|_t,\{\de\theta^c\}_{-\infty}^{t-\de t})\right]^2}{r\left(\Delta\theta,\{\de\theta^c\}_{-\infty}^{t-\de t}\right)} \right\rangle + \mathcal{O}\left(\epsilon^3\right).
\end{equation}
In this section, we use angle brackets to denote averages over the full distribution $p\left(\{\theta\}_{-\infty}^t,\{\de\theta^c\}_{-\infty}^t\right)$.
The leading-order term is zero if and only if $\eta = \bar\eta$, so that the optimal control strategy cannot depend on memory of past states in the low-information limit.

We then note that since the optimal distribution in the small-$\epsilon$ limit differs from uniform only in the first Fourier mode, the control protocol should likewise depend only on the first Fourier mode to first order. Any dependence on higher Fourier modes will increase the information rate without improving performance. While dependence on both $\sin\theta$ and $\cos\theta$ can both be used to control the first Fourier mode, only $\cos\theta$ satisfies the constraint that only $|\theta|$ can be used. Thus we can write
\begin{equation}
\lambda\left(\Delta\theta,|\theta|,\{\de\theta^c\}_{-\infty}^{t-\de t}\right) = r\left(\Delta\theta|\{\de\theta^c\}_{-\infty}^{t-\de t}\right) + \epsilon\, A\left(\Delta\theta,\{\de\theta^c\}_{-\infty}^{t-\de t}\right)\cos\theta + \mathcal{O}\left(\epsilon^2\right).
\end{equation}
We now impose stationarity of the steady-state distribution, or equivalently of the first Fourier mode in the small-$\epsilon$ limit. Passive diffusion decreases the mode $\cos\theta$ at a rate $\epsilon D_r/\sqrt{2}$, so the active control must increase the first mode at the same rate. A jump by $\Delta\theta$ increases the first mode $\cos\theta$ by
\begin{equation}
\begin{aligned}
\Delta\cos\theta & = \cos(\theta + \Delta\theta) - \cos\theta\\
& = -\left[1-\cos\Delta\theta\right]\cos\theta - \sin\theta\sin\Delta\theta\\
& = -z(\Delta\theta)\cos\theta - \sin\theta\sin\Delta\theta,
\end{aligned}
\end{equation}
where we have defined the function $z(\Delta\theta)\equiv1-\cos\Delta\theta$. The active control therefore increases the first mode at a rate
\begin{equation}\label{eq:steadystatecondition}
\begin{aligned}
&  \left\langle \int\de\Delta\theta\,\lambda\left(\Delta\theta,|\theta|,\{\de\theta^c\}_{-\infty}^{t-\de t}\right)\left[-z(\Delta\theta)\cos\theta - \sin\theta\sin\Delta\theta\right]\right\rangle\\
& = -\frac{\epsilon}{\sqrt{2}}\left\langle \int\de\Delta\theta\,r\left(\Delta\theta|\{\de\theta^c\}_{-\infty}^{t-\de t}\right)z(\Delta\theta)\right\rangle - \frac{\epsilon}{2}\left\langle \int\de\Delta\theta\,A\left(\Delta\theta,\{\de\theta^c\}_{-\infty}^{t-\de t}\right)z(\Delta\theta)\right\rangle + \mathcal{O}\left(\epsilon^2\right)\\
& = +\frac{\epsilon D_r}{\sqrt{2}}+ \mathcal{O}\left(\epsilon^2\right).
\end{aligned}
\end{equation}
Here in the last line we used the fact that to first order in $\epsilon$ this must be equal to the passive diffusion rate $-\epsilon D_r/\sqrt{2}$.

We can now directly evaluate the transfer entropy rate, which to second order in $\epsilon$ is
\begin{equation}
\begin{aligned}
\dot{\mathcal{T}}_{\theta\to\de\theta^c} &= \frac{\epsilon^2}{2}\left\langle\int\de\Delta\theta \frac{\left[ A\left(\Delta\theta,\{\de\theta^c\}_{-\infty}^{t-\de t}\right)\cos\theta\right]^2}{r\left(\Delta\theta|\{\de\theta^c\}_{-\infty}^{t-\de t}\right)}\right\rangle + \mathcal{O}\left(\epsilon^3\right)\\
& = \frac{\epsilon^2}{4}\left\langle\int\de\Delta\theta \frac{A\left(\Delta\theta,\{\de\theta^c\}_{-\infty}^{t-\de t}\right)^2}{r\left(\Delta\theta|\{\de\theta^c\}_{-\infty}^{t-\de t}\right)}\right\rangle + \mathcal{O}\left(\epsilon^3\right).
\end{aligned}
\end{equation}
We then apply the Cauchy--Schwarz inequality, taking $X=A/\sqrt{r}$ and $Y=\sqrt{r}z$ such that $\langle X^2\rangle\geq |\langle XY\rangle|^2/\langle Y^2\rangle$. Combined with Eq.~\eqref{eq:steadystatecondition}, this gives
\begin{equation}\label{eq:SItransferentropybound}
\begin{aligned}
\dot{\mathcal{T}}_{\theta\to\de\theta^c} &\geq \frac{\epsilon^2}{4} \frac{\left\langle \int\de\Delta\theta\,A\left(\Delta\theta,\{\de\theta^c\}_{-\infty}^{t-\de t}\right)z(\Delta\theta)\right\rangle^2}{\left\langle \int\de\Delta\theta\,r\left(\Delta\theta|\{\de\theta^c\}_{-\infty}^{t-\de t}\right)z(\Delta\theta)^2\right\rangle} + \mathcal{O}\left(\epsilon^3\right)\\
& = \frac{\epsilon^2}{2} \frac{\left(\left\langle \int\de\Delta\theta\,r\left(\Delta\theta|\{\de\theta^c\}_{-\infty}^{t-\de t}\right)z(\Delta\theta)\right\rangle + D_r\right)^2}{\left\langle \int\de\Delta\theta\,r\left(\Delta\theta|\{\de\theta^c\}_{-\infty}^{t-\de t}\right)z(\Delta\theta)^2\right\rangle} + \mathcal{O}\left(\epsilon^3\right).
\end{aligned}
\end{equation}
Recalling that $z(\Delta\theta)=1-\cos\Delta\theta$, we note that $z(\Delta\theta)\leq2$, so that we can write 
\begin{equation}
\left\langle \int\de\Delta\theta\,r\left(\Delta\theta|\{\de\theta^c\}_{-\infty}^{t-\de t}\right)z(\Delta\theta)^2\right\rangle \leq 2 \left\langle \int\de\Delta\theta\,r\left(\Delta\theta|\{\de\theta^c\}_{-\infty}^{t-\de t}\right)z(\Delta\theta)\right\rangle.
\end{equation}
Using this inequality in Eq.~\eqref{eq:SItransferentropybound} then gives
\begin{equation}\label{eq:SItransferentropyinequality2}
\begin{aligned}
\dot{\mathcal{T}}_{\theta\to\de\theta^c} & \geq \frac{\epsilon^2}{4} \frac{\left(\left\langle \int\de\Delta\theta\,r\left(\Delta\theta|\{\de\theta^c\}_{-\infty}^{t-\de t}\right)z(\Delta\theta)\right\rangle + D_r\right)^2}{\left\langle \int\de\Delta\theta\,r\left(\Delta\theta|\{\de\theta^c\}_{-\infty}^{t-\de t}\right)z(\Delta\theta)\right\rangle} + \mathcal{O}\left(\epsilon^3\right)\\
& = \epsilon^2D_r + \frac{\epsilon^2}{4} \frac{\left(\left\langle \int\de\Delta\theta\,r\left(\Delta\theta|\{\de\theta^c\}_{-\infty}^{t-\de t}\right)z(\Delta\theta)\right\rangle - D_r\right)^2}{\left\langle \int\de\Delta\theta\,r\left(\Delta\theta|\{\de\theta^c\}_{-\infty}^{t-\de t}\right)z(\Delta\theta)\right\rangle} + \mathcal{O}\left(\epsilon^3\right).
\end{aligned}
\end{equation}
Consistency with our definition $\epsilon\equiv \left(\dot{\mathcal{T}}_{\theta\to\de\theta^c}/D_r\right)^{1/2}\ll1$ requires $\dot{\mathcal{T}}_{\theta\to\de\theta^c} = \epsilon^2 D_r$ to second order. Thus saturation of the Pareto frontier in the low-information limit requires both saturating the inequality~\eqref{eq:SItransferentropyinequality2}, and
\begin{equation}\label{eq:SIthirdcondition}
\frac{\left(\left\langle \int\de\Delta\theta\,r\left(\Delta\theta|\{\de\theta^c\}_{-\infty}^{t-\de t}\right)z(\Delta\theta)\right\rangle - D_r\right)^2}{\left\langle \int\de\Delta\theta\,r\left(\Delta\theta|\{\de\theta^c\}_{-\infty}^{t-\de t}\right)z(\Delta\theta)\right\rangle} = 0.
\end{equation}
The inequality, Eq.~\eqref{eq:SItransferentropyinequality2}, came from applying both the Cauchy--Schwarz inequality as well as $z(\Delta\theta)\leq 2$. For the second inequality, saturation is only possible for $\Delta\theta=\pi$. Saturating Cauchy--Schwarz requires $A/\sqrt{r}\propto \sqrt{r}z$, which with $z(\pi)=2$ implies $A\left(\Delta\theta,\{\de\theta^c\}_{-\infty}^{t-\de t}\right) \propto r\left(\Delta\theta|\{\de\theta^c\}_{-\infty}^{t-\de t}\right)$. The remaining condition, Eq.~\eqref{eq:SIthirdcondition}, requires
\begin{equation}
\left\langle \int\de\Delta\theta\,r\left(\Delta\theta|\{\de\theta^c\}_{-\infty}^{t-\de t}\right)z(\Delta\theta)\right\rangle = D_r.
\end{equation}
With only jumps $\Delta\theta=\pi$ and $z(\pi)=2$, this gives $\langle r\left(\pi|\{\de\theta^c\}_{-\infty}^{t-\de t}\right)\rangle = D_r/2$. Eq.~\eqref{eq:steadystatecondition} then implies $A\left(\pi,\{\de\theta^c\}_{-\infty}^{t-\de t}\right) = -2\sqrt{2} r\left(\pi|\{\de\theta^c\}_{-\infty}^{t-\de t}\right)$.

Combining these results, the control protocol that saturates the Pareto frontier to first order in $\epsilon$ is
\begin{equation}
\lambda\left(\Delta\theta|\{|\theta|\}_{-\infty}^{t},\{\de\theta^c\}_{-\infty}^{t-\de t}\right) = r\left(\pi|\{\de\theta^c\}_{-\infty}^{t-\de t}\right)\left[1 - \epsilon
\,2\sqrt{2}\cos\theta\right]\delta(\Delta\theta-\pi) + \mathcal{O}\left(\epsilon^2\right).
\end{equation}
This proves that, subject to the constraint of only observing $|\theta|$, the optimal protocol in the low-information limit uses no memory of past observations, and only induces discrete jumps by $\Delta\theta=\pi$. This is a classical Run-Reverse strategy.

The remaining degeneracy allows for arbitrary dependence of the jump rate on the control history $\{\de\theta^c\}_{-\infty}^{t-\de t}$. A specific choice with no control-history dependence would be $r\left(\pi|\{\de\theta^c\}_{-\infty}^{t-\de t}\right) = \left\langle r\left(\pi|\{\de\theta^c\}_{-\infty}^{t-\de t}\right)\right\rangle$, which leads to the control protocol 
\begin{equation}
\lambda\left(\Delta\theta|\{|\theta|\}_{-\infty}^t\right) = \frac{D_r}{2}\left[1 - \epsilon
\,2\sqrt{2}\cos\theta\right]\delta(\Delta\theta-\pi) + \mathcal{O}\left(\epsilon^2\right),
\end{equation}
which exactly recovers the optimal reverse rate function derived in Ref.~\cite{betancourt2026discrete}.

%%%%%%%%%%%%%%%%%%%%%%%%%%%%%%%%%%%%%%%%%%%%%%%%%%%%%%%%%%%%%%
\section{Details for Information Engine}\label{app:infoengine}
Here we provide full details for the information engine application example in the main text. There we consider the colloidal information engine which has been implemented experimentally in Ref.~\cite{saha2021maximizing}. The controlled system is a Brownian particle of mass $m$ with vertical position $x$, diffusing in a harmonic potential with center $z$. The overdamped Langevin dynamics are
\begin{equation}
\de x_t = \beta D\kappa (z_t-x_t)\de t - \beta Dmg\de t + \sqrt{2 D}\de W^p_t,
\end{equation}
with control update $\de z_t = \de z^c_t$. Here $D$ is the particle's diffusion coefficient, $\kappa$ is the trap strength, $g$ is the gravitational acceleration, and $\beta$ is the inverse temperature. The controller moves the trap center $z$, and we require that the controller do no work on the particle on average, i.e., that the controller does not change the potential energy $V(x,z) = \frac{1}{2}\kappa(z-x)^2$. Performance is quantified by the rate, $\mathcal{P}[\rho] = \beta\dot{Q}$, at which heat is extracted from the bath, equal to the rate of free energy storage as gravitational potential energy. To apply the bound~\eqref{eq:main_inequality}, consider the difference coordinate $u=z-x$ which reaches a steady state $\rho(u)$.

\subsection{Deriving the Pareto Frontier}
Here we derive the Pareto frontier for the information engine, Eq.~\eqref{eq:info_engine_frontier}. Applying the bound~\eqref{eq:main_inequality} to the information engine, and evaluating the passive entropy rate, gives (using $u=z-x$, and the steady-state probability $\rho(u)$):
\begin{equation}\label{eq:SI_infoeng_bound}
D\mathcal{I}[\rho] - \beta D\kappa\leq \dot{\mathcal{T}}_{u\to\de z^c}.
\end{equation}
We define the timescale $\tau_r=1/\beta D\kappa$ and dimensionless mass $\delta_g = mg/\kappa\sqrt{D\tau_r}$, as well as the length scale $\sigma = \sqrt{D\tau_r}$. We can write the steady-state upward velocity of the particle as 
\begin{equation}
v\tau_r/\sigma = \langle u\rangle/\sigma - \delta_g.
\end{equation}

To do no work on the particle on average, the controller must not change the potential energy $V(x,z) = \frac{1}{2}\kappa(x-z)^2$. Equivalently, the control updates must not change $\langle u^2\rangle$. It follows that at steady state the passive contribution to the rate of change of $\langle u^2\rangle$ must also vanish. This results in a one-to-one relationship between the variance of $u$ and the velocity: $\mathrm{Var}(u)/\sigma^2 = 1 - \delta_g \nu - \nu^2$, where we have defined the dimensionless velocity $\nu\equiv v\tau_r/\sigma$. 

The Cramér--Rao inequality gives $\mathcal{I}[\rho]\geq 1/\mathrm{Var}(u)$, which, combined with the bound~\eqref{eq:SI_infoeng_bound}, gives
\begin{equation}
\frac{1}{1-\delta_g\nu - \nu^2} - 1\leq \tau_r\dot{\mathcal{T}}_{u\to\de z^c}.
\end{equation}
Solving for $\nu$ yields a velocity--information frontier:
\begin{equation}
\nu\leq \frac{-\delta_g + \sqrt{\delta_g^2 + 4/\left(1+\left(\tau_r\dot{\mathcal{T}}_{u\to\de z^c}\right)^{-1}\right)}}{2}.
\end{equation}
The velocity limit increases monotonically with information, asymptotically approaching the finite upper bound $\frac{1}{2} \left(\sqrt{\delta_g^2 + 4}-\delta_g\right)$ as $\dot{\mathcal{T}}_{u\to\de z^c}\to\infty$. In this way, achieving the maximum performance of the engine requires a diverging information rate. Multiplying by $\delta_g$ leads to the Pareto frontier in the main text between extracted heat and information rate, Eq.~\eqref{eq:info_engine_frontier}. Optimizing over $\delta_g$ gives a more general bound,
\begin{equation}
\tau_r\beta\dot{Q}\leq \frac{1}{1 + \left(\tau_r\dot{\mathcal{T}}_{u\to \de z^c}\right)^{-1}}.
\end{equation}

\subsection{Second Law for the Information Engine}
Here we show that the second law implies $\beta\dot{Q}\leq \dot{\mathcal{T}}_{u\to\de z^c}$. Let $x$ and $z$ be bipartite degrees of freedom, as described in Ref.~\cite{ehrich2023energy}. Then the second law of bipartite thermodynamics applied to the $x$ subsystem gives~\cite{horowitz2014thermodynamics,ehrich2023energy}
\begin{equation}
\beta\dot{Q}\leq \dot{I}_Z \equiv \lim_{\de t\to0}\frac{I[x_t;z_{t+\de t}]-I[x_t;z_t]}{\de t},
\end{equation}
for information flow $\dot{I}_Z$ defined as the instantaneous rate at which the $z$ dynamics increase the mutual information between $x$ and $z$. This is a lower bound on the transfer entropy rate from $x$ to $z$, giving~\cite{horowitz2014second}
\begin{equation}
\dot{I}_Z\leq \dot{\mathcal{T}}_{x\to z} \equiv \lim_{\de t\to0}\frac{1}{\de t}I[z_{t+\de t};\{x\}_{-\infty}^t|\{z\}_{-\infty}^t].
\end{equation}
Finally, for the particular information engine setup, the transfer entropy rate $\dot{\mathcal{T}}_{x\to z}$ is equal to the control information rate $\dot{\mathcal{T}}_{u\to\de z^c}$. To see this, note that the dynamics depend only on the state $u=z-x$, not the absolute values of $z$ or $x$, so that
\begin{equation}
\begin{aligned}
\dot{\mathcal{T}}_{x\to z} & \equiv \lim_{\de t\to0}\frac{1}{\de t}I[z_{t+\de t};\{x\}_{-\infty}^t|\{z\}_{-\infty}^t]\\
& = \lim_{\de t\to0}\frac{1}{\de t}I[\de z_{t}^c;\{x\}_{-\infty}^t|\{z\}_{-\infty}^t]\\
& = \lim_{\de t\to0}\frac{1}{\de t}I[\de z_{t}^c;\{u\}_{-\infty}^t|\{z\}_{-\infty}^t]\\
& = \lim_{\de t\to0}\frac{1}{\de t}I[\de z_{t}^c;\{u\}_{-\infty}^t|\{\de z^c\}_{-\infty}^{t-\de t}]\\
& = \dot{\mathcal{T}}_{u\to\de z^c}.
\end{aligned}
\end{equation}
Thus the second law implies $\beta\dot{Q}\leq \dot{\mathcal{T}}_{u\to\de z^c}$.

\subsection{Experimentally Implemented Control Protocol}
The controller makes noisy measurements of the particle position $\hat{x}_t = x_t + \mathcal{N}(0,\sigma_m^2)$ (with measurement noise $\sigma_m$) at periodic time intervals of length $t_s$. The protocol implemented in Ref.~\cite{saha2021maximizing} updates the trap center based only on the measured particle position as
\begin{equation}\label{eq:trap_update}
\de z^c_{t = nt_s} = \begin{cases} \alpha(\hat x_t-z_t), & \hat{x}_t>z_t,\\
0, & \mathrm{otherwise},\end{cases}
\end{equation}
for integer $n$, and feedback gain parameter $\alpha$ that is selected to ensure the controller does no average work on the particle. For simplicity, we omit the feedback delay in Ref.~\cite{saha2021maximizing}; this would be a useful direction for future work.

\subsection{Probabilistic Time-Reversal Protocol}
Here we construct the time-reversal control protocol for the information engine shown in Fig.~\ref{fig:infoengine}. We will initially neglect the measurement noise, then restore it at the end of this subsection.

For notational simplicity, we use the dimensionless variable $\xi = (z-x)/\sigma$, which obeys dynamics
\begin{equation}
\de \xi_\tau = (\delta_g-\xi)\de \tau + \sqrt{2}\de W^p_\tau + \de\xi^c_\tau.
\end{equation}
Here $\tau = t/\tau_r$ is scaled time. The controller acts at discrete time intervals of scaled time $\Delta = t_s/\tau_r$. We will use $\xi^-$ and $\xi^+$ for the states immediately before and after a control update respectively. The control action is thus $\de\xi^c = \xi^+-\xi^- = \mathrm{d}z^c/\sigma$.

Saturating the Pareto frontier requires a Gaussian distribution (to saturate the Cramér--Rao inequality), so we take $\xi^-$ and $\xi^+$ to be Gaussian-distributed with means $m_\pm$ and variances $s_\pm$, as
\begin{subequations}
\begin{align}
\xi^- \sim\mathcal{N}(m_-,s_-),\\
\xi^+ \sim \mathcal{N}(m_+,s_+).
\end{align}
\end{subequations}
The mean velocity of the particle in scaled units is thus $\nu = (m_+-m_-)/\Delta$. Now consider the passive kernel from $\xi^+$ to $\xi^-$ over time $\Delta$,
\begin{equation}
K_p(\xi^-|\xi^+) = \mathcal{N}\left[\xi^+ + (\delta_g - \xi^+)(1-e^{-\Delta}), 1-e^{-2\Delta} \right].
\end{equation}
This implies
\begin{subequations}
\begin{align}
m_- & = m_+ + (\delta_g - m_+)(1-e^{-\Delta}),\\
s_- & = s_+ e^{-2\Delta} + 1-e^{-2\Delta}.
\end{align}
\end{subequations}
Requiring that the trap does no average work on the particle enforces $\langle(\xi^-)^2\rangle=\langle(\xi^+)^2\rangle$, which in turn implies
\begin{equation}
m_-^2 + s_- = m_+^2 + s_+.
\end{equation}
Thus we can write the two means and variances in terms of only the parameters $\Delta$ and $\delta_g$, as well as the dimensionless velocity $\nu$, as
\begin{subequations}
\begin{align}
m_+
&=
\delta_g
+
\frac{\Delta \nu}{1-e^{-\Delta}},
\\
m_-
&=
\delta_g
+
\frac{\Delta \nu}{e^{\Delta}-1},
\\
s_+
&=
1
-
\left(
\frac{\Delta \nu}{1-e^{-\Delta}}
\right)^2
-
\frac{2\delta_g \Delta \nu}{1-e^{-2\Delta}},
\\
s_-
&=
1
-
\left(
\frac{\Delta \nu}{e^{\Delta}-1}
\right)^2
-
\frac{2\delta_g \Delta \nu}{e^{2\Delta}-1}.
\end{align}
\end{subequations}

The reverse passive kernel, meanwhile, is
\begin{equation}
R_p(\xi^+|\xi^-) = \frac{p_+(\xi^+)K_p(\xi^-|\xi^+)}{p_-(\xi^-)} = \mathcal{N}\left[m^+ + \frac{s_+}{s_-}e^{-\Delta}(\xi^--m_-) , \frac{s_+}{s_-}(1-e^{-2\Delta})\right].
\end{equation}
Thus to probabilistically time-reverse the passive dynamics, the control kernel must be
\begin{equation}
K_c(\xi^+|\xi^-) = R_p(\xi^+|\xi^-).
\end{equation}
We can write the control, at time $\tau=n\Delta$ ($n\in\mathbb{N}$), in the differential notation used throughout the \type as
\begin{equation}\label{eq:infoenginetimereversalprotocol}
\de\xi^c_\tau(\xi_\tau^-) = (m_+-m_-) + \left(\frac{s_+}{s_-}e^{-\Delta}-1\right)(\xi^-_\tau - m_-) + \sqrt{\frac{s^+}{s_-}(1-e^{-2\Delta})}\,\mathcal{N}(0,1).
\end{equation}

We can now straightforwardly account for the measurement noise, by noting that the dimensionless pre-control measurement will be $\hat{\xi}^-_\tau = \xi^-_\tau + \mathcal{N}(0,\sigma_m^2/\sigma^2)$. We can then achieve the probabilistic time-reversal protocol, Eq.~\eqref{eq:infoenginetimereversalprotocol}, using the noisy measurement $\hat{\xi}^-_\tau$, by simply decreasing the noise added to the control update. The protocol, as a function of the noisy measurement, will be
\begin{equation}
\de\xi^c_\tau(\hat{\xi}_\tau^-) = (m_+-m_-) + \left(\frac{s_+}{s_-}e^{-\Delta}-1\right)(\hat{\xi}^-_\tau - m_-) + \sqrt{\frac{s^+}{s_-}(1-e^{-2\Delta}) - \left(\frac{s_+}{s_-}e^{-\Delta}-1\right)^2\frac{\sigma^2_m}{\sigma^2}}\,\mathcal{N}(0,1),
\end{equation}
which leads to a distribution $p(\de\xi_\tau^c|\xi_\tau^-)$ identical to the target time-reversal control distribution with no measurement noise. This is achievable so long as the measurement noise satisfies the condition
\begin{equation}
\frac{s^+}{s_-}(1-e^{-2\Delta}) - \left(\frac{s_+}{s_-}e^{-\Delta}-1\right)^2\frac{\sigma^2_m}{\sigma^2}\geq 0.
\end{equation}

\subsection{Information Rate for the Time-Reversal Protocol}
For the time-reversal protocol, we can compute the transfer entropy rate semi-analytically. For a single controller update, the transfer entropy is
\begin{equation}
\begin{aligned}\label{eq:infoenginetimereversalterate}
\dot{\mathcal{T}}_{\xi\to\de \xi^c} & = \frac{1}{\Delta} I\left[\de \xi^c_\tau;\{\xi\}_{-\infty}^\tau|\{\de \xi^c\}_{-\infty}^{\tau-\Delta}\right]\\
& = \frac{1}{\Delta}I\left[\de \xi^c_\tau;\xi_\tau^-|\{\de \xi^c\}_{-\infty}^{\tau-\Delta}\right]\\
& = \frac{1}{\Delta} \left(S\left[\de \xi^c_\tau|\{\de \xi^c\}_{-\infty}^{\tau-\Delta}\right] - S\left[\de \xi^c_\tau|\xi_\tau^-,\{\de \xi^c\}_{-\infty}^{\tau-\Delta}\right] \right).
\end{aligned}
\end{equation}
Here in the second line we used the fact that the time-reversal control protocol depends only on the current pre-control state $\xi_\tau^-$. The remaining task is then to evaluate the two conditional entropy terms. Given knowledge of $\xi_\tau^-$ the only remaining variability in the controller update is the Gaussian white noise, so the second conditional entropy is
\begin{equation}
S\left[\de \xi^c_\tau|\xi_\tau^-,\{\de \xi^c\}_{-\infty}^{\tau-\Delta}\right] = \frac{1}{2}\ln\left(2\pi e \frac{s^+}{s_-}(1-e^{-2\Delta})\right).
\end{equation}
The first conditional entropy also requires the variance $\mathrm{Var}(\de\xi^c_\tau|\{\de\xi^c\}_{-\infty}^{\tau-\Delta})$. To compute this variance, we consider how it changes due to controlled and passive updates. Using Eq.~\eqref{eq:infoenginetimereversalprotocol}, the variance of the control update will be
\begin{equation}\label{SIeq:var1}
\mathrm{Var}(\de\xi^c_\tau|\{\de\xi^c\}_{-\infty}^{\tau-\Delta}) = \left(\frac{s_+}{s_-}e^{-\Delta}-1\right)^2 \mathrm{Var}(\xi_\tau^-|\{\de\xi^c\}_{-\infty}^{\tau-\Delta}) + \frac{s^+}{s_-}(1-e^{-2\Delta}).
\end{equation}
We can then write, noting that $\xi_\tau^-$ and $\de\xi^c_\tau$ are jointly Gaussian distributed and using the standard equation for the conditional variance of a joint,
\begin{equation}\label{SIeq:var2}
\mathrm{Var}(\xi^-_\tau|\de\xi^c_\tau,\{\de\xi^c\}_{-\infty}^{\tau-\Delta}) = \mathrm{Var}(\xi_\tau^-|\{\de\xi^c\}_{-\infty}^{\tau-\Delta}) - \frac{\mathrm{Cov}(\de\xi^c_\tau, \xi_\tau^-|\{\de\xi^c\}_{-\infty}^{\tau-\Delta})^2}{\mathrm{Var}(\de\xi^c_\tau|\{\de\xi^c\}_{-\infty}^{\tau-\Delta})}.
\end{equation}
Here the conditional covariance evaluates to $\mathrm{Cov}(\de\xi^c_\tau, \xi_\tau^-|\{\de\xi^c\}_{-\infty}^{\tau-\Delta}) = \left(\frac{s_+}{s_-}e^{-\Delta}-1\right)\mathrm{Var}(\xi_\tau^-|\{\de\xi^c\}_{-\infty}^{\tau-\Delta})$.

Since $\xi^+_\tau$ is a deterministic translation of $\xi^-_\tau$ given $\de\xi^c_\tau$, we also have $\mathrm{Var}(\xi^+_{\tau}|\{\de\xi^c\}_{-\infty}^{\tau}) = \mathrm{Var}(\xi^-_{\tau}|\{\de\xi^c\}_{-\infty}^{\tau})$. The passive update from time $\tau$ to $\tau+\Delta$ then leads to pre-control variance conditional on the control history at time $\tau+\Delta$ given by
\begin{equation}\label{SIeq:var3}
\mathrm{Var}(\xi^-_{\tau+\Delta}|\{\de\xi^c\}_{-\infty}^{\tau}) = \mathrm{Var}(\xi^+_\tau|\{\de\xi^c\}_{-\infty}^{\tau})e^{-2\Delta} + 1-e^{-2\Delta}.
\end{equation}
Using the fact that $\mathrm{Var}(\xi^-_{\tau+\Delta}|\{\de\xi^c\}_{-\infty}^{\tau})=\mathrm{Var}(\xi_\tau^-|\{\de\xi^c\}_{-\infty}^{\tau-\Delta})$ at steady state, Eqs.~\eqref{SIeq:var1}, \eqref{SIeq:var2}, and \eqref{SIeq:var3} then constitute three coupled equations in three unknown variables. We then numerically solve for the variance $\mathrm{Var}(\de\xi^c_\tau|\{\de\xi^c\}_{-\infty}^{\tau-\Delta})$ of the next control update given the control history in terms of only the parameters $\Delta$, $\delta_g$, and $\nu$. The transfer entropy rate is then computed using Eq.~\eqref{eq:infoenginetimereversalterate}. Returning to physical units then simply requires dividing by the dimensional timescale $\tau_r$.

%%%%%%%%%%%%%%%%%%%%%%%%%%%%%%%%%%%%%%%%%%%%%%%
\section{Numerical Transfer Entropy Rate Calculations}
When analytic calculations are not tractable, we numerically compute the information rate $\dot{\mathcal{T}}_{\x\to\de\x^c}$ using the Path Weight Sampling method (TE-PWS) developed in Ref.~\cite{das2025exact}. This approach is used for the time-reversal protocol in Fig.~\ref{fig:mexicanhat}, the Run-Reverse navigation strategy in Fig.~\ref{fig:navigation}, and the experimental information engine protocol in Fig.~\ref{fig:infoengine}. Our code for implementing the TE-PWS method, adapted from code made public as part of Ref.~\cite{das2025exact}, is available on GitHub~\cite{github}.

%\newpage
%%%%%%%%%%%%%%%%%%%%%%%%%%%%%%%%%%%%%%%%%%%%%%%%
\bibliography{main}% Produces the bibliography via BibTeX.